\documentclass[useAMS,usenatbib]{mnras}

\usepackage{graphicx}
\usepackage{color}
\usepackage{multirow}
\usepackage{amssymb,amsmath}

\def\aap{A\&A}
\def\apj{ApJ}

\def\apjl{ApJL}
\def\aj{AJ}
\def\pasp{PASP}
\def\pasa{PASA}
\def\mnras{MNRAS}
\def\nat{Nature}
\def\na{New Astron.}

\def\apjs{ApJS}

\def\ps{$\rm km \,s^{-1}\,kpc^{-1}$}

\def\kms{\,km\,s$^{-1}$}

\def\H2{$\rm H_2$}

\title[Tidal barred-spirals in different galaxy models]{Bars and spirals in tidal interactions with an ensemble of galaxy mass models}

\author[A. R. Pettitt \& J. W. Wadsley]{Alex R. Pettitt$^{1}$\thanks{E-mail:
alex@astro1.sci.hokudai.ac.jp} and J. W. Wadsley$^2$\\
$^{1}$Department of Physics, Faculty of Science, Hokkaido University, Sapporo 060-0810, Japan\\
$^{2}$Department of Physics and Astronomy, McMaster University, Hamilton, Canada\\
}

\begin{document}

\date{\today}

\pagerange{\pageref{firstpage}--\pageref{lastpage}} \pubyear{0000}

\maketitle

\label{firstpage}

\begin{abstract}
We present simulations of the gaseous and stellar material in several different galaxy mass models under the influence of different tidal fly-bys to assess the changes in their bar and spiral morphology. Five different mass models are chosen to represent the variety of rotation curves seen in nature. We find a multitude of different spiral and bar structures can be created, with their properties dependent on the strength of the interaction. We calculate pattern speeds, spiral wind-up rates, bar lengths, and angular momentum exchange to quantify the changes in disc morphology in each scenario. The wind-up rates of the tidal spirals follow the 2:1 resonance very closely for the flat and dark matter dominated rotation curves, whereas the more baryon dominated curves tend to wind-up faster, influenced by their inner bars. Clear spurs are seen in most of the tidal spirals, most noticeable in the flat rotation curve models. Bars formed both in isolation and interactions agree well with those seen in real galaxies, with a mixture of ``fast" and ``slow" rotators. We find no strong correlation between bar length or pattern speed and the interaction strength. Bar formation is, however, accelerated/induced in four out of five of our models. We close by briefly comparing the morphology of our models to real galaxies, easily finding analogues for nearly all simulations presenter here, showing passages of small companions can easily reproduce an ensemble of observed morphologies.
\end{abstract}


%%%%%%%%%%%%%%%%%%%%%%%%%%%%%%%%%%%%%%%%%%%%%%%%
%%%%%%%%%%%%%%%%%%%%%%%%%%%%%%%%%%%%%%%%%%%%%%%%
%%%%%%%%%%%%%%%%%%%%%%%%%%%%%%%%%%%%%%%%%%%%%%%%
\section{Introduction}
%%%%%%%%%%%%%%%%%%%%%%%%%%%%%%%%%%%%%%%%%%%%%%%%
%%%%%%%%%%%%%%%%%%%%%%%%%%%%%%%%%%%%%%%%%%%%%%%%
%%%%%%%%%%%%%%%%%%%%%%%%%%%%%%%%%%%%%%%%%%%%%%%%

Disc galaxies are known to display a wide variety of structures in both their stellar and gaseous components \citep{1936rene.book.....H,1959HDP....53..275D,2008MNRAS.389.1179L,2011A&A...532A..74B}. The most prominent of these features are the striking inner bars and spiral arms, with some galaxies, such as our own Milky Way, believed to harbour both. Quite how these structures are generated, and what maintains them, has been the subject of many decades of observations and theoretical studies.

Interactions between galaxies are believed to be commonplace, be they between similar sized galactic discs, dwarfs, or dark matter subhaloes \citep{1984ApJ...278L..71S}. They can induce changes in the star formation properties of the galaxies \citep{1978ApJ...219...46L,1985AJ.....90..708K,1987AJ.....93.1011K} and many examples exist of galaxies that appear in mid-interaction, be they early stages of mergers \citep{2000AJ....120..630E,2013MNRAS.430.1901H,2015MNRAS.446.2038R} or fly-by encounters \citep{1999IAUS..186...81Y,2005MNRAS.364...69S,2016A&A...588A..33Q}. Cosmological simulations suggest that interactions are an essential part of a galaxy's history (e.g. \citealt{2010MNRAS.404..575L,2014ApJ...780...57B,2015MNRAS.452.4347K}), and thus play an important part in the evolution of spiral and bar features.

The classical picture of spiral arms is that they exist as density wave like feature, with gas and stars flowing through the spiral pattern \citep{1964ApJ...140..646L,1973PASAu...2..174K}. Gas and dust lanes are seen to trace these spiral arm features \citep{2003PASP..115..928K,2008AJ....136.2563W,2011ApJ...730...72R}, with gas believed to experience strong shocks as it falls into the spiral potential well \citep{1968IAUS...29..453F,1969ApJ...158..123R}. While the details of this theory have changed somewhat over the years (e.g. \citealt{1989ApJ...338...78B,1996ssgd.book.....B}) the keystone idea has remained the same until somewhat recently. Results of numerical simulations have fuelled the theory of dynamical spiral arms, where arms are recurrent transients that rotate with the material speed of the disc, winding up as they do so \citep{1984ApJ...282...61S,1993A&A...272...37E}, rather than at a near-fixed pattern speed like the classical density wave picture. These two different theories seem at odds in a number of respects, including the prevalence of star-gas arm offsets, arm lifetimes and locations of shocks \citep{2010MNRAS.409..396D,2011ApJ...735....1W,2012MNRAS.426..167G,2015MNRAS.453.1867G,2015PASJ...67L...4B}. We refer the reader to recent reviews on the subject for details \citet{2011MNRAS.410.1637S} and \citet{2014PASA...31...35D}.

Spiral arms can be readily induced in galaxy interactions \citep{1972ApJ...178..623T,1991A&A...252..571D,2010MNRAS.403..625D}, creating M51-like tidal bridges and tails. The shape of the resulting spirals depend on the orbital paths and masses of the galactic components \citep{1991A&A...244...52E,2015ApJ...807...73O,2016MNRAS.458.3990P}. Simulations of such interactions suggest they also induce strong bursts of star formation in the tidal arms \citep{1986MNRAS.219..305N,2007A&A...468...61D,2016A&A...592A..62G,2017MNRAS.468.4189P}. These spirals behave somewhat differently to density wave and dynamic spiral patterns, in that they are semi-long lived, outlasting dynamic spirals but still winding up over time, with a pattern speed much slower than the material rotation speed.

Bars are seen in a number of disc galaxies, with fractions ranging from a quarter to nearly half of observed disc galaxies, depending on the classification and sample selection \citep{2009A&A...495..491A,2011MNRAS.411.2026M}. They can be formed readily in isolation in kinematically cold and massive enough stellar discs \citep{1971ApJ...168..343H,1973ApJ...186..467O}, emerging in a fast growth phase due to an initial instability, then later buckling out of the disc plane \citep{1990A&A...233...82C,1991Natur.352..411R}. Later they enter a secular evolution phase, gradually growing and slowing down \citep{2013seg..book....1K,2013seg..book..305A,2014RvMP...86....1S}. The impact of bars on the internal properties of galaxies have been studied by many simulations and observations in the past (e.g. \citealt{2001PASJ...53.1163W,2005ApJ...632..217S,2013MNRAS.436.1836R,2015MNRAS.453.3082P}), including their effect on molecular clouds properties, migrational infall, and the orbits of stars and gas. 

While there have been many studies focused on the formation and properties of bars in isolation, comparatively few have focussed on how tidal interactions impact bars. Early numerical work suggested the clear-cut picture that interactions with galactic discs will reduce bar formation times and induce bar features in discs that were bar-free in isolation \citep{1987MNRAS.228..635N}. The bars that are formed are seen to show little difference to their isolated counterparts, with pattern speeds and morphologies being determined foremost by the rotation curve of the host galaxy \citep{1990A&A...230...37G,1991A&A...243..118S}. Conversely, later work suggested that bars formed in interactions rotate slower than those formed in isolation \citep{1998ApJ...499..149M,2014MNRAS.445.1339L,2017MNRAS.464.1502M}. The picture became further muddled by studies showing certain interactions can impede/cease bar formation rather than induce it \citep{2003IAUS..208..177A,2017A&A...604A..75M,2017arXiv170502348Z}, with discs sensitive to a given orbital path and mass ratio to induce bar growth \citep{2014ApJ...790L..33L,2016MNRAS.463.2210K,2017ApJ...842...56G}. Studies have also looked into interactions other than fly-bys, such as the growth of bars within dwarfs around some host halo \citep{2001ApJ...559..754M,2006A&A...447..453C,2008ApJ...687L..13R}. There is growing observational evidence that the presence of bars in discs may be environmentally dependent \citep{2012ApJ...761L...6M,2012MNRAS.423.1485S}, with dense environments having a greater bar-hosting population. These dense environments could imply a higher rate of interactions and/or higher relative speeds, which imply bar features are in some way dependent on interactions occurring in the wider environment.

The picture of quite how interactions impact bar and spiral properties is not complete, with studies usually confined to looking at a particular feature in a specific galaxy model, or with insufficient numerical resolution to properly capture the dynamics of the stellar (and gaseous) disc. The aim of this work is to study the response of a variety of disc galaxies to perturbing satellite passages, specifically focussing on bar and spiral features, via a suite of $N$-body and hydrodynamical simulations. For instance, how are bar lengths and pattern speeds changed in a tidal interaction, and how long-lived are tidal spiral arms for galaxies with varying levels of shear?

This paper is organised as follows. Section \ref{sec:numerics} details the computational method and the generation of the initial conditions for both the galaxies and the interaction scenarios. Results are presented and discussed in Section \ref{sec:results}, where we discuss the general morphology, pattern speeds, spiral arms, bars, and observational analogues. We then conclude in Section \ref{sec:conc}.

%%%%%%%%%%%%%%%%%%%%%%%%%%%%%%%%%%%%%%%%%%%%%%%%
%%%%%%%%%%%%%%%%%%%%%%%%%%%%%%%%%%%%%%%%%%%%%%%%
%%%%%%%%%%%%%%%%%%%%%%%%%%%%%%%%%%%%%%%%%%%%%%%%
\section{Numerical Simulations}
\label{sec:numerics}
%%%%%%%%%%%%%%%%%%%%%%%%%%%%%%%%%%%%%%%%%%%%%%%%
%%%%%%%%%%%%%%%%%%%%%%%%%%%%%%%%%%%%%%%%%%%%%%%%
%%%%%%%%%%%%%%%%%%%%%%%%%%%%%%%%%%%%%%%%%%%%%%%%

%%%%%%%%%%%%%%%%%%%%%%%%%%%%%%%%%%%%%%%%%%%%%%%%
%%%%%%%%%%%%%%%%%%%%%%%%%%%%%%%%%%%%%%%%%%%%%%%%
\subsection{Numerics}
%%%%%%%%%%%%%%%%%%%%%%%%%%%%%%%%%%%%%%%%%%%%%%%%
%%%%%%%%%%%%%%%%%%%%%%%%%%%%%%%%%%%%%%%%%%%%%%%%

Simulations were performed using the $N$-body, smoothed particle hydrodynamics (SPH) code \textsc{gasoline2} \citep{2004NewA....9..137W,2017arXiv170703824W}. Gravity is solved using a binary tree, and the system integrated using a kick-drift-kick leapfrog. We use 64 neighbours and the standard cubic spline kernel. Self-gravity is active for all components, using a fixed gravitational softening of 50pc. The gas is isothermal with a temperature of 10,000\,K, in effect simulating the warm interstellar medium (ISM) and halting gravitational collapse of the gas. The effects of cooling, and resulting star formation/feedback processes in tidal spirals are not included here, though were the focus of \citet{2017MNRAS.468.4189P}, which we refer the reader to for an in depth discussion of the multi-phase ISM and star forming properties of such discs.

%%%%%%%%%%%%%%%%%%%%%%%%%%%%%%%%%%%%%%%%%%%%%%%%
%%%%%%%%%%%%%%%%%%%%%%%%%%%%%%%%%%%%%%%%%%%%%%%%
\subsection{Galaxy models}
%%%%%%%%%%%%%%%%%%%%%%%%%%%%%%%%%%%%%%%%%%%%%%%%
%%%%%%%%%%%%%%%%%%%%%%%%%%%%%%%%%%%%%%%%%%%%%%%%
 
\begin{figure}
\includegraphics[trim = 10mm 10mm 0mm 0mm,width=90mm]{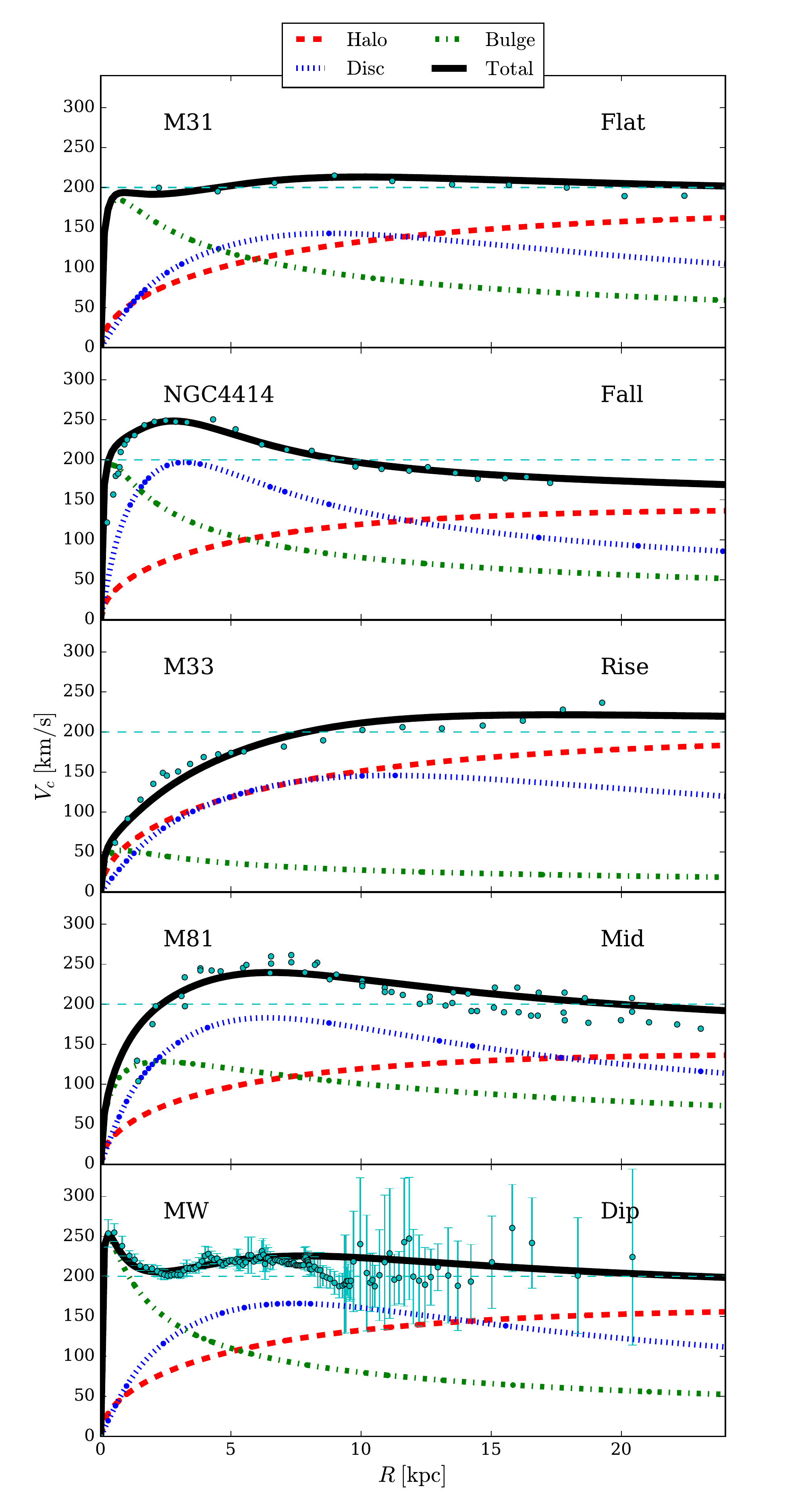}
 \caption{The five different galactic rotation curves under investigation. Each is composed of a disc, bulge and halo component of differing masses and scale lengths tailored to match the general structure of six different observed galaxies. Sources of rotation curve data are given in the main text.}
\label{RcSp}
\end{figure}

We choose to setup galaxies using the \textsc{galic} initial conditions generator \citep{2014MNRAS.444...62Y} where the galaxy is decomposed into a exponential disc\footnote{In setting up our interacting simulations we made use of the Miyamoto-Nagai decomposition approximation of \citet{2015MNRAS.448.2934S} to express the three dimensional potential resulting from the disc density profile $\rho(r,z)\propto \exp{(-r/r_d)}{\rm sech}(-|z|/z_d)$ in an analytic form.}, and Hernquist profile bulge and halo. We use 1 million particles for the gas disc, 1\;million for stellar disc, 1\;million for the dark halo, 50,000 for the stellar bulge, and 10,000 for the companions.
The angular momentum transfer between the halo and disc particles plays an important part in the formation of the bar. Thus, using a live halo is preferential to using a static potential to properly encapsulate the growth of possible bars. \citet{2009ApJ...697..293D} study the impact of different resolutions of halo and disc components on bar structures. As resolution increases, there is a noticeable change in the bar formation time, and thus the resulting phase angle, though resolution tends to have a lesser impact on the fundamental properties such as pattern speed and bar length (see also \citealt{2014MNRAS.444...62Y,2017RMxAA..53..257V}); so long as haloes are resolved with approximately a million mass elements. While there is some evidence that haloes need to be resolved with more than $\mathcal{O}(10^8)$ particles to fully encapsulate their long term evolution \citep{2007MNRAS.375..425W}, this is deemed excessive by follow-up studies, suggesting a more manageable $\mathcal{O}(10^5)$ -- $\mathcal{O}(10^6)$ is sufficient \citep{2008ApJ...679..379S}. Regarding spiral arms: \citet{2011ApJ...730..109F} found that 1 million disc particles are sufficient to resolve dynamic, $N$-body spiral arms in simulations for many Gyr. In light of these works, our adopted disc resolution should be sufficient to capture spiral arm and bar structure in both the isolated and interacting systems.

Figure\;\ref{RcSp} shows the five different rotation curves for the galaxy models presented here, with relevant parameters given in Table\;\ref{Models}. The general shape of each rotation curve is chosen to match certain shapes of rotation curves seen in nature. These include rotation curves that are flat (standard baryon/dark matter configuration), falling (bulge+disc dominated, minimal dark halo component), rising (dark matter dominated), peaked in the mid-disc (disc dominated), and dipped due to a strong central bulge concentration. Rotation curves for each model are based on the following: M31 \citep{2003ApJ...588..311W},  NGC\,4414 \citep{1998ASPC..136..193B}, M33 \citep{2000MNRAS.311..441C}, M81 \citep{2014ApJ...785..103F} and the Milky Way \citep{2012PASJ...64...75S}, shown as cyan points in Figure\;\ref{RcSp}. Models are tailored by matching the masses of the bulge, disc and halo ($M_b$, $M_d$, $M_h$) and their respective scale-lengths ($a_b$, $a_d$, $a_h$). Each rotation curve has been re-normalised to ensure that the mid-disc velocity is approximately 200\kms{}, which is justified in that we are only interested in the shape and relative mass contributions in the disc rather than the absolute values. In many cases choosing the halo/disc/bulge parameters was somewhat degenerate, in which case components were tailored to ensure a varied sample among the models and in turn the resulting spiral/bar morphology. We stress that these models are not meant to re-create the morphologies of each specific observed galaxy, instead they serve simply to provide a varied ensemble of rotation curve types as seen in nature.

\begin{table*}
\centering
 \begin{tabular}{@{}l c c c c c c c c c c}
  \hline
  Model & $M_d$ & $a_d$  & $M_b$ & $a_b$ & $M_h$ & $a_h$ & $\bar{Q}_s$& $\epsilon_{\rm bar}$ & $m_{\rm swing} (R=2a_d)$ & $f_{\rm obs}$\\
  \hline 
  \hline
  Flat & 5.14 & 3.90 & 2.06 & 0.64 & 114.21 & 42.92 & 1.3 & 0.90 & 3.4 & 0.4 \\
  Fall & 3.57 & 1.43 & 1.53 & 0.44 & 51.07 & 29.26   & 1.6 & 0.80 & 2.2 & 0.1 \\
  Rise & 6.78 & 4.94 & 0.20 & 0.81 & 135.57 & 40.52  & 1.4  & 0.90 & 3.2 & 0.3 \\
  Mid & 6.13 & 2.85 & 3.57 & 2.34 & 51.07 & 29.26  & 1.5  & 0.75 & 2.2 & 0.1 \\
  Dip & 5.88 & 3.29 & 1.57 & 0.27 & 78.45 & 33.77  & 1.3  & 0.94 & 2.7 & 0.1\\
 \end{tabular}
 \caption{Parameters used for each mass model depicted in Figure\;\ref{RcSp}. Masses and scale length are given in units of $10^{10}M_\odot$ and $\rm kpc$ respectively. $\bar{Q}_s$ is the mean Toomre-$Q$ parameter in the stars between $a_d$ and $2a_d$. The swing amplified spiral mode is given in the mid/outer disc region; $R=2a_d$. The $f_{\rm obs}$ parameter indicates the approximate fraction of galaxies from \citet{2016PASJ...68....2S} that correspond to this a given rotation curve type.}
 \label{Models}
\end{table*}

\begin{figure}
\includegraphics[trim = 0mm 0mm -10mm 0mm,width=90mm]{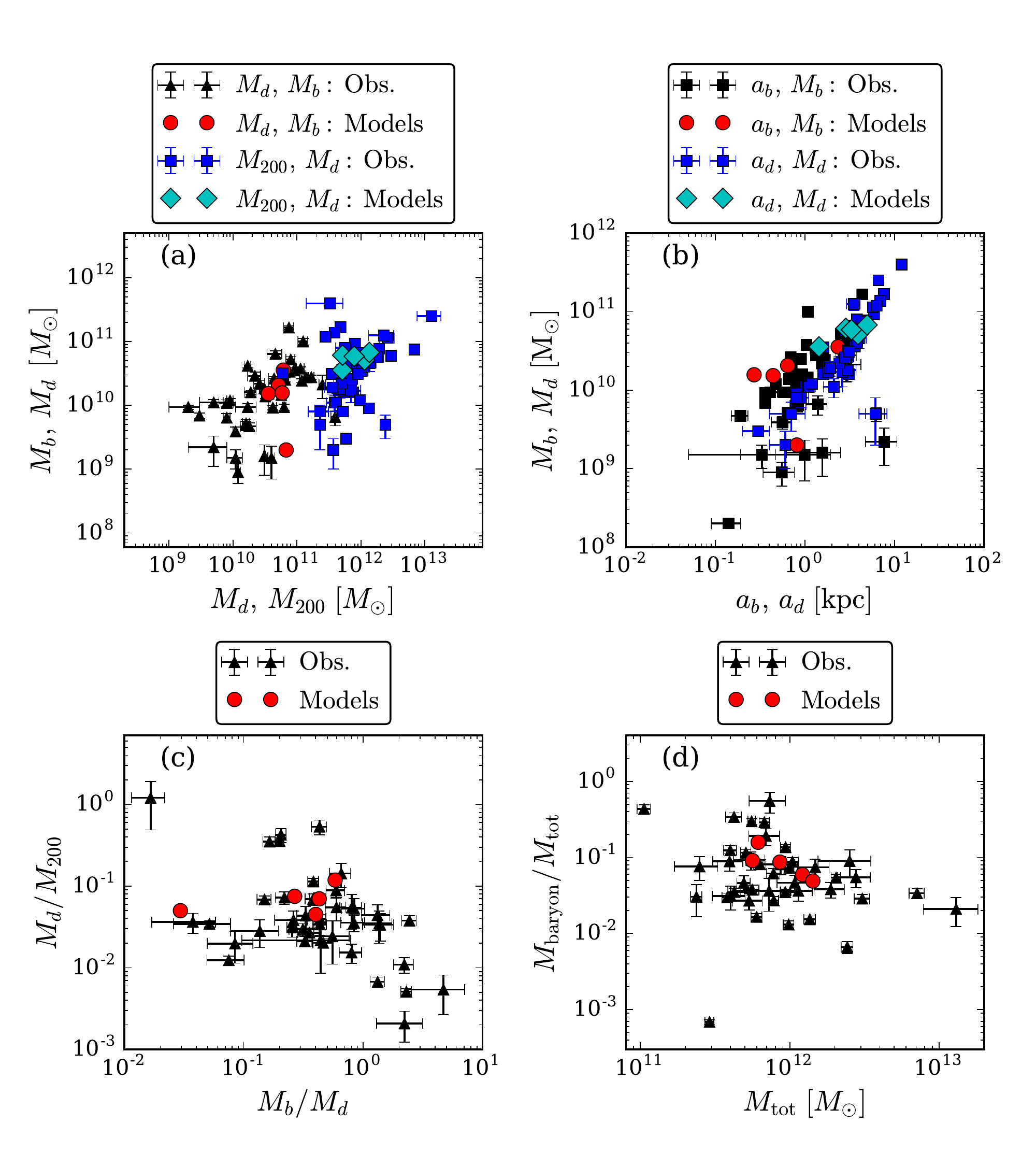}
   \caption{Model parameters plotted against observed values for 43 galaxies from \citet{2016PASJ...68....2S}. Observed galaxies are shown by black and blue points with errorbars, and the five models used in this study as red circles and cyan diamonds. The different panels show: (a) mass-mass relations; $M_d$:$M_b$ and $M_{200}$:$M_d$, (b) mass-size relation; $a_b$:$M_b$ and $a_d$:$M_d$, (c) bulge to disc and disc to halo mass ratios, and (d) total mass ratios, where $M_{\rm baryon}$ includes disc and bulge components, and $M_{\rm tot}$ includes disc, bulge and dark halo.}
\label{obsgalvalues}
\end{figure}

In Table\;\ref{Models} we give an approximate value for the ratio of observed galaxies with each type of rotation curve, $f_{\rm obs}$. This is calculated from the rotation curve catalogue of \citet{2016PASJ...68....2S}, which contains over 100 well-resolved rotation curves compiled from the literature. Galaxy rotation curves were simply tagged as one of the five types by-eye. While this is a crude estimator and does not account for any inherent bias in the sample selection, it does gives a good general picture of the rotation curves seen in nature. Note that this contains no correction for type, with certain galaxy types favouring certain rotation curves (e.g. rising curves are abundant in dwarfs; \citealt{2009A&A...493..871S}). To further justify our choice of models, we plot the various mass and scale lengths in Figure\;\ref{obsgalvalues} against observed scale lengths and masses presented in \citet{2016PASJ...68....2S}. Black and blue points with errorbars show values from observed galaxies, with red and cyan points showing our adopted models. Note that our models occupy quite a small region of parameter space in some panels due to the re-normalisation of the rotation curves to 200\kms{}. In all cases our models lie within the parameter space seen in observed galaxies, with the only minor difference being a tendency for a higher bulge mass in our models for a given bulge scale length compared to observations. However, 1:1 correspondence can hardly be expected due to differences between the mass profiles used between this study and \citet{2016PASJ...68....2S} for each component.

For all calculations we add an exponential gas disc with a mass of $M_g=0.1M_d$. The gas disc scale length is set as $a_g=2a_d$, and a maximum disc extent of double that of the stellar disc to match the general trend see in observed galaxies (e.g. \citealt{1997A&A...324..877B}). Gas positions are copies of disc stars particles, but with their radii doubled and velocities re-drawn from the rotation curve plus a dispersion of 10\kms{}. In doing so the galaxy is now somewhat out of the equilibrium state produced by the \textsc{galic} code. As such, once the gas disc is included we employ the azimuthal shuffling technique of \citet{2007MNRAS.378..541M} to ensure a radial equilibrium state before time integration. We shuffle particles every 1\,Myr for a period of 1\,Gyr which allowed radial oscillations to dissipate in all of the rotation curves investigated.

The Toomre parameter in the stars, $Q_s$ \citep{1964ApJ...139.1217T}, characterises the stability of the disc to local collapse and is given by
\begin{equation}
Q_s = \frac{\kappa \sigma_R}{3.36 G \Sigma_d}
\end{equation}
where $\kappa$ is the epicycle frequency, $\sigma_R$ the radial velocity dispersion and $\Sigma_d$ the disc surface density. Values of $Q_s<1$ imply the disc is gravitationally unstable. The dominant swing amplified mode of the stellar disc, denoted $m_{\rm swing}$ \citep{1981seng.proc..111T}, can be calculated by
\begin{equation}
m_{\rm swing}=\frac{\kappa^2 R}{2\pi G \Sigma_d X}\approx \frac{\kappa^2 R}{4\pi G \Sigma_d}
\label{SwingAmp}
\end{equation}
where $1<X< 2$ generates spiral features, and $X=2$ is a nominally adopted value \citep{2011ApJ...730..109F,2014PASA...31...35D}. These models are also capable of forming bar-structures on somewhat longer time-scales. A metric for bar formation is given by \citet{1982MNRAS.199.1069E} as;
\begin{equation}
\epsilon_{\rm bar}=\frac{V_{\rm max}}{\sqrt{GM_d/a_d}}
\end{equation}
where $V_{\rm max}$ is the maximum disc rotation velocity, with discs being unstable to bar-formation if $\epsilon_{\rm bar}<1.1$. We note that the latter criterion for the stability of discs to form bars is quite a crude estimator, and several studies have shown that kinematically hot discs embedded in relatively massive dark haloes are still able to form bars \citep{2002ApJ...569L..83A,2013MNRAS.434.1287S,2016ApJ...819...92S}. We use this parameter as an approximate proxy for bar instability with these caveats in mind.

\begin{figure}
\includegraphics[trim = 0mm 0mm 0mm 0mm,width=80mm]{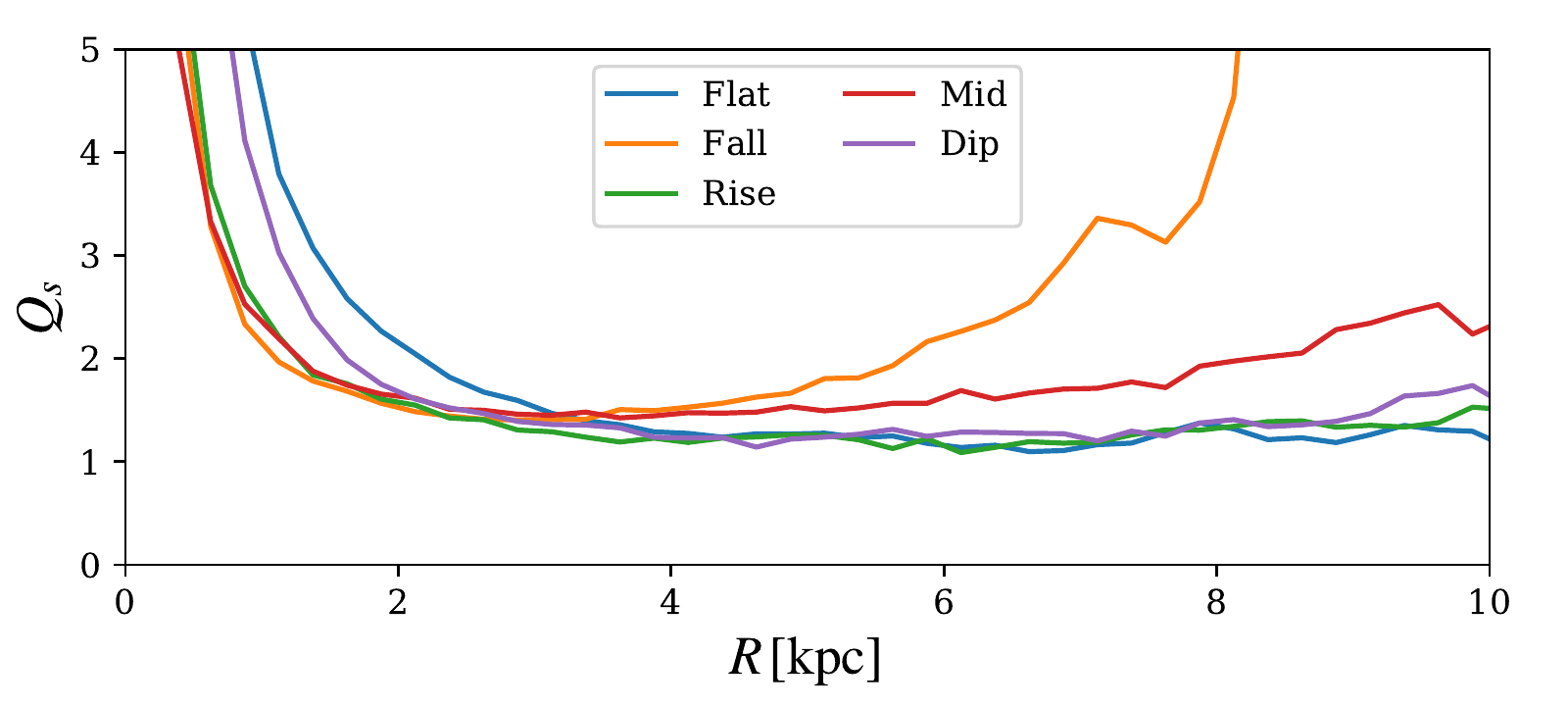}
 \caption{Toomre-$Q$ parameter in the stars at initialisation after the azimuthal shuffling period for each of the five rotation curves.}
\label{FigToomre}
\end{figure}

These three parameters give insight on the general structure formed in disc galaxies. $Q_s$ indicates if the disc is sufficiently gravitationally dominated against kinematic support to experience some fragmentation. $m_{\rm swing}$ indicates the number of arms is a strong function of the disc mass, with systems with high disc-to-halo mass ratios forming only a few strong spiral arms, whereas low mass discs form numerous but weaker arms \citep{1985ApJ...298..486C,2015MNRAS.449.3911P,2015ApJ...808L...8D}. Finally $\epsilon_{\rm bar}$ suggests bar formation is a strong function of the balance between disc and halo masses (the latter determining the magnitude of $V_{\rm max}$). Each model is tailored so that the Toomre-$Q$ parameter in the stars is near unity at initialisation. The shuffling process increases $Q_s$ slightly over the course of the 1\,Gyr integration, bringing $Q_s$ to 1.3--1.6 depending on the model (see Table\;\ref{Models}). Figure\;\ref{FigToomre} shows $Q_s$ for each model at the beginning of the simulation. Each model shows similar trends, with $1.<Q_s<1.6$ in the main disc region. $Q_s$ rises sharply in the inner disc, and the outer disc for Fall model in particular, though there is very little disc material at these larger radii. Table\,\ref{Models} also gives values for $\epsilon_{\rm bar}$ and $m_{\rm swing}$ at initialisation (after the shuffling epoch). All discs appear somewhat bar unstable, with the Fall and Mid discs especially so, implying they will form bars much faster than, for example, Dip. Despite the $\epsilon_{\rm bar}$ parameter indicating a bar unstable nature, Flat and Dip proved to be quite stable to bar features in early preliminary simulations. This may be due to the gas disc acting as a stabilising factor which is not taken into account by $\epsilon_{\rm bar}$. Flat and Dip therefore act as our bar triggering case studies. Mid and Fall are expected to form bars in isolation, so they are our focus for identifying how interactions alter the bar dynamics and morphology. While tidal triggering of bars is of interest, it is not our sole aim, as such we chose a mix of discs that appear highly unstable to bar formation and some that appear borderline stable over moderate time-scales. As for spiral structure, Mid and Fall are predicted to form the lowest arm numbers and to display 2-armed spirals in the mid disc, with Flat appearing the most flocculent and generating something between a 3--4 armed spiral.

%%%%%%%%%%%%%%%%%%%%%%%%%%%%%%%%%%%%%%%%%%%%%%%%
%%%%%%%%%%%%%%%%%%%%%%%%%%%%%%%%%%%%%%%%%%%%%%%%
\subsection{Interaction model}
%%%%%%%%%%%%%%%%%%%%%%%%%%%%%%%%%%%%%%%%%%%%%%%%
%%%%%%%%%%%%%%%%%%%%%%%%%%%%%%%%%%%%%%%%%%%%%%%%

The strength of the galaxy--companion interaction can be characterised by the dimensionless parameter \citep{1991A&A...244...52E}:
\begin{equation}
S =  \left(\frac{R_{\rm enc}}{d} \right)^3  \frac{\Delta T}{T}\frac{M_p}{M_{\rm tot}(R<R_{\rm enc})},
\label{Seq}
\end{equation}
where $M_p$ is the companion mass and $d$ is the distance of closest approach. $R_{\rm enc}$ is a characteristic distance of the galaxy (taken here to be 20kpc, the truncation distance of the stellar disc) and $M_{\rm tot}(R<R_{\rm enc})$ is the mass of all the host galaxy components within this radius. $\Delta T$ is the time for the perturber to move 1 radian at closest approach, and $T$ is the time for stars at $R_{\rm enc}$ to move 1 radian in orbit around the galactic centre. This $S$ parameter provides information on the tidal strength of the interaction and is the force experienced by material in the outer edge of the disc over a duration $\Delta T$ as a fraction of the circular momentum in the galactic orbit at this point.

\begin{table}
\centering
 \begin{tabular}{@{}l c c c  }
  \hline
  Model & $S$ & $M_p$  \\
  \hline 
  \hline
  FlatS10 & 0.10 & 5.00 \\
  FlatS05 & 0.05 &  2.50     \\
  FlatS00 & - & 0.00 \\
  \hline
  FallS10 & 0.10 & 3.00  \\
  FallS05 & 0.05 & 1.50      \\
  FallS00 & - & 0.00      \\
  \hline
  RiseS10  & 0.10 & 6.00       \\
  RiseS05  & 0.05 & 3.00       \\
  RiseS00 & - & 0.00    \\
  \hline
  MidS10  & 0.10 & 3.70     \\
  MidS05  & 0.05 &  1.85      \\
  MidS00 & - & 0.00 \\
 \hline
  DipS10 & 0.10 &  4.10      \\
  DipS05 & 0.05 &   2.05     \\
  DipS00 & - & 0.00  \\
 \end{tabular}
 \caption{Perturber parameters for each calculation included here. We uniformly define closest approach as $b=20$kpc. Masses are given in units of $10^{10}M_\odot$. The strength parameter, $S$, is defined by Eq.\,\ref{Seq}.}
 \label{Perturbers}
\end{table}

We use the $S$ parameter to constrain the strength of each interaction for each galactic disc. All our orbits are parabolic, prograde and with a fixed closest approach distance (20kpc). We choose two different interaction strengths; $S=0.1$ and $0.05$ which represent a strong and weak interaction scenario. While studies find that values of $S$ in the range of $0.01\leq S \leq 0.25$ can produce some sort of arm response, values below $S<0.05$ produce only a very weak spiral structure visible in the outer disc or for a very short duration, with values higher than 0.25 causing widespread disruption of the primary's disc structure \citep{1991A&A...244...52E,2008ApJ...683...94O,2016MNRAS.458.3990P}. Determining $S$ from observations is somewhat complicated, due to the impulse dependence, though minimum values for some interacting galaxies have been calculated. \citet{1995AJ....110.2605M} found $S_{\rm min}=0.04$ for an disc-elliptical pair in the Hickson Compact Groups HCG 47a, and \citet{1995ApJ...453..139E} found the IC\,2163+NGC\,2207 system well fit by a model with $S=0.12$. As the closest approach is fixed, only the mass of perturber need be changed to alter the strength of the interaction for unbound orbital paths. Table\,\ref{Perturbers} gives the value of the mass for the companion needed for a given interaction strength for each galaxy model (listing all 15 models discussed in this work). It is noted that strongly bound orbits tend to create more diverse structures (e.g. M51 or NGC\,4676) but we are interested in unbound orbits where the origin of galactic structure is less clear.

The companion is introduced into a simulation after 0.4\,Gyr of isolated evolution. Closest approach occurs approximately 0.2\,Gyr later, and we run each simulation for a total time of 2.4\,Gyr (except for the Dip runs, where we run for a total of 6\,Gyr to follow the later emergence of a bar feature). This allows us to follow the evolution for 1.8\,Gyr after companion passage, or approximately 7--9 galactic rotations.

Our adopted configurations give initial relative velocities of the companion--host system of 230--340\kms{}, depending on the host galaxy mass model (Fall being the lowest, Rise the highest). This then rises to values of 320--450\kms{} at perigalacticon passage, falling off again as the companion moves away. Such relative velocities are not dissimilar to galaxy clusters and groups. These range from a couple of hundred to over a thousand \kms{} \citep{1993ApJ...404...38G,1999ApJS..125...35S}, with the Local Group displaying a mean velocity dispersion of only 62\kms{} \citep{1994ApJ...436...23B}. While the relative velocities for our models are noticeably higher than the rotational velocities at closest approach (thereby not consistent with the quasi-resonances described by \citealt{2010ApJ...725..353D}), they are nearly exactly twice the rotational velocity of each galaxy at this radii, suggesting a 2:1 resonance effect may be playing a part in driving tidal features.

%%%%%%%%%%%%%%%%%%%%%%%%%%%%%%%%%%%%%%%%%%%%%%%%
%%%%%%%%%%%%%%%%%%%%%%%%%%%%%%%%%%%%%%%%%%%%%%%%
%%%%%%%%%%%%%%%%%%%%%%%%%%%%%%%%%%%%%%%%%%%%%%%%
\section{Results and discussion}
\label{sec:results}
%%%%%%%%%%%%%%%%%%%%%%%%%%%%%%%%%%%%%%%%%%%%%%%%
%%%%%%%%%%%%%%%%%%%%%%%%%%%%%%%%%%%%%%%%%%%%%%%%
%%%%%%%%%%%%%%%%%%%%%%%%%%%%%%%%%%%%%%%%%%%%%%%%

%%%%%%%%%%%%%%%%%%%%%%%%%%%%%%%%%%%%%%%%%%%%%%%%
%%%%%%%%%%%%%%%%%%%%%%%%%%%%%%%%%%%%%%%%%%%%%%%%
\subsection{General morphology}
\label{sec_morph}
%%%%%%%%%%%%%%%%%%%%%%%%%%%%%%%%%%%%%%%%%%%%%%%%
%%%%%%%%%%%%%%%%%%%%%%%%%%%%%%%%%%%%%%%%%%%%%%%%

%%%%%%%%%%%%%%%%%%%%%%%%%%%%%%%%%%%%%%%%%%%%%%%%
\subsubsection{Disc structure}
%%%%%%%%%%%%%%%%%%%%%%%%%%%%%%%%%%%%%%%%%%%%%%%%
\begin{figure*}
\centering
\resizebox{.90\hsize}{!}{\includegraphics[trim = 0mm 0mm 0mm 0mm]{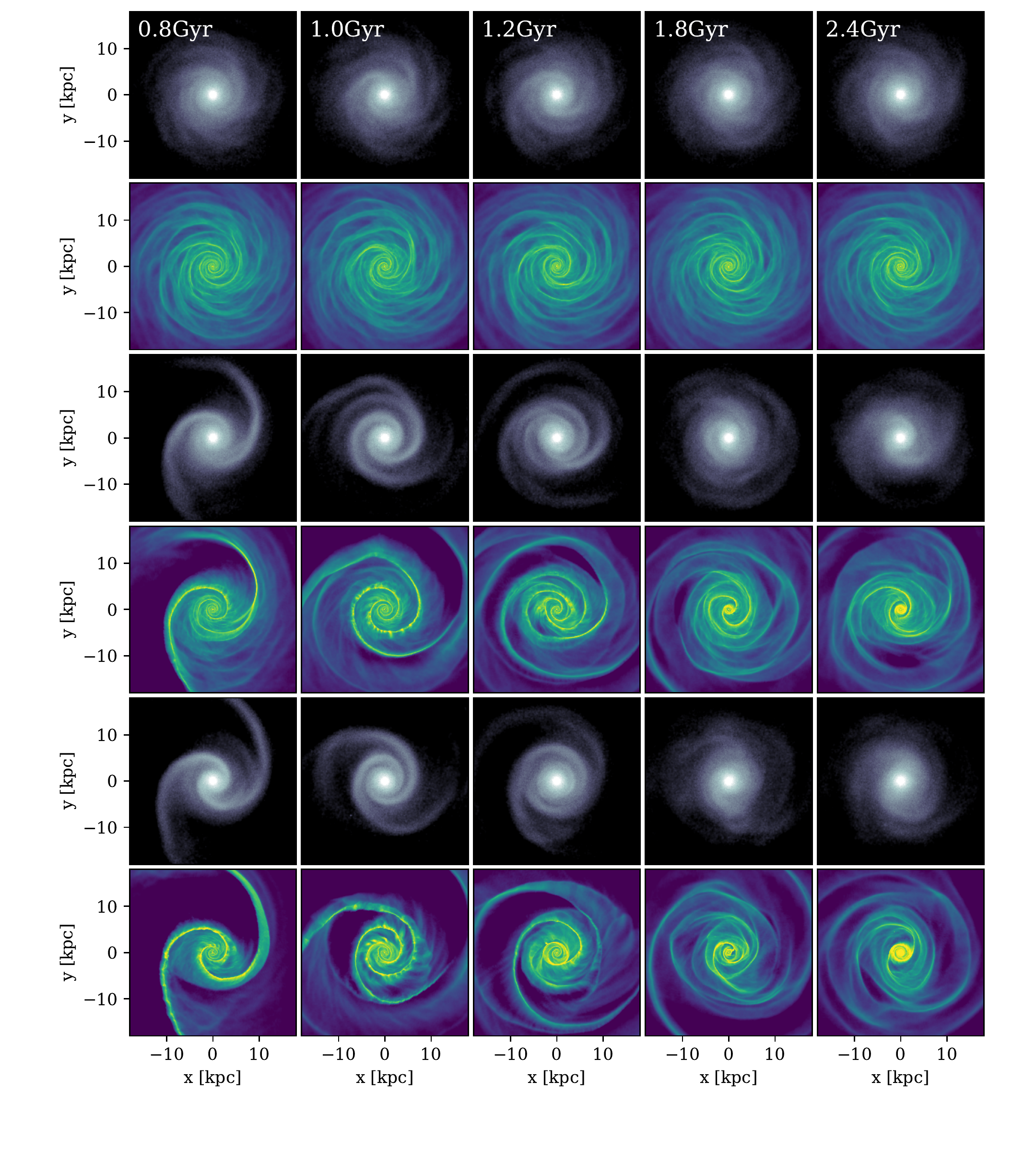}}
\caption{Face-on evolution of all three calculations for the Flat model. The top two rows show the isolated system (stars above the gas), the middle rows show the S05 perturbed model, and the bottoms rows the S10 model.}
\label{IsoGalsA}
\end{figure*}

\begin{figure*}
\centering
\resizebox{.90\hsize}{!}{\includegraphics[trim = 0mm 0mm 0mm 0mm]{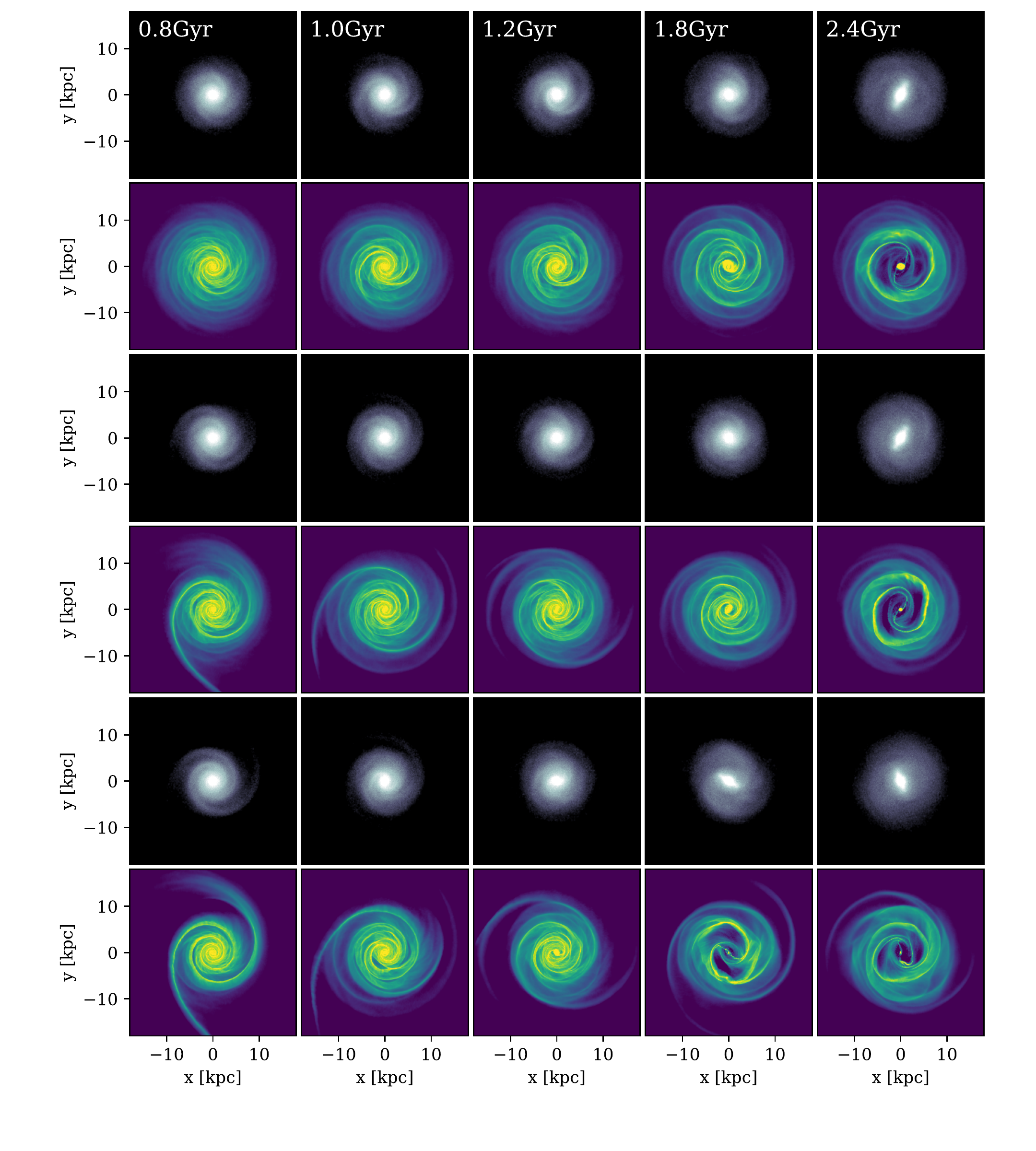}}
\caption{As Fig. \ref{IsoGalsA} but for the Fall calculations.}
\label{IsoGalsB}
\end{figure*}

\begin{figure*}
\centering
\resizebox{.90\hsize}{!}{\includegraphics[trim = 0mm 0mm 0mm 0mm]{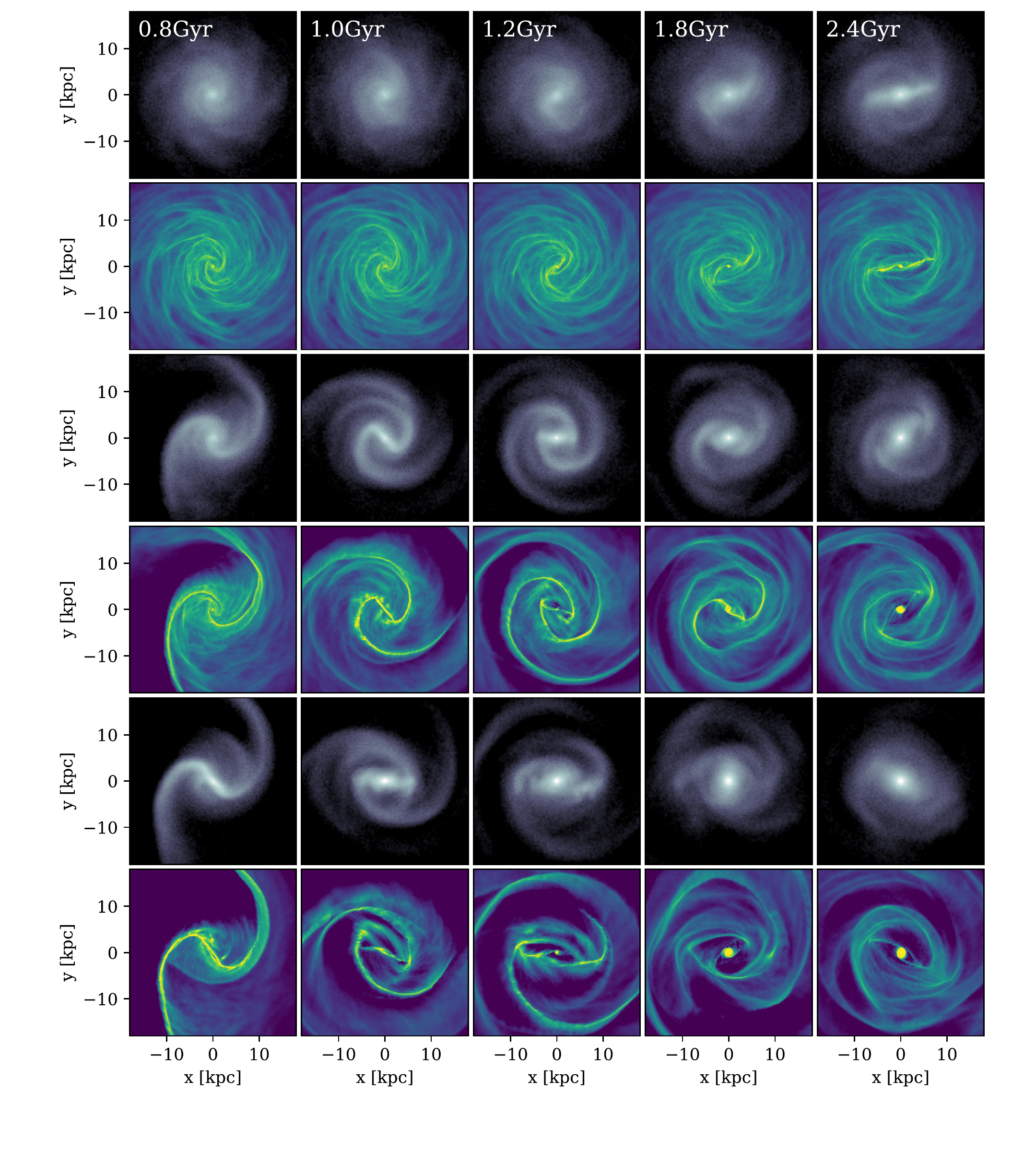}}
\caption{As Fig. \ref{IsoGalsA} but for the Rise calculations.}
\label{IsoGalsC}
\end{figure*}

\begin{figure*}
\centering
\resizebox{.90\hsize}{!}{\includegraphics[trim = 0mm 0mm 0mm 0mm]{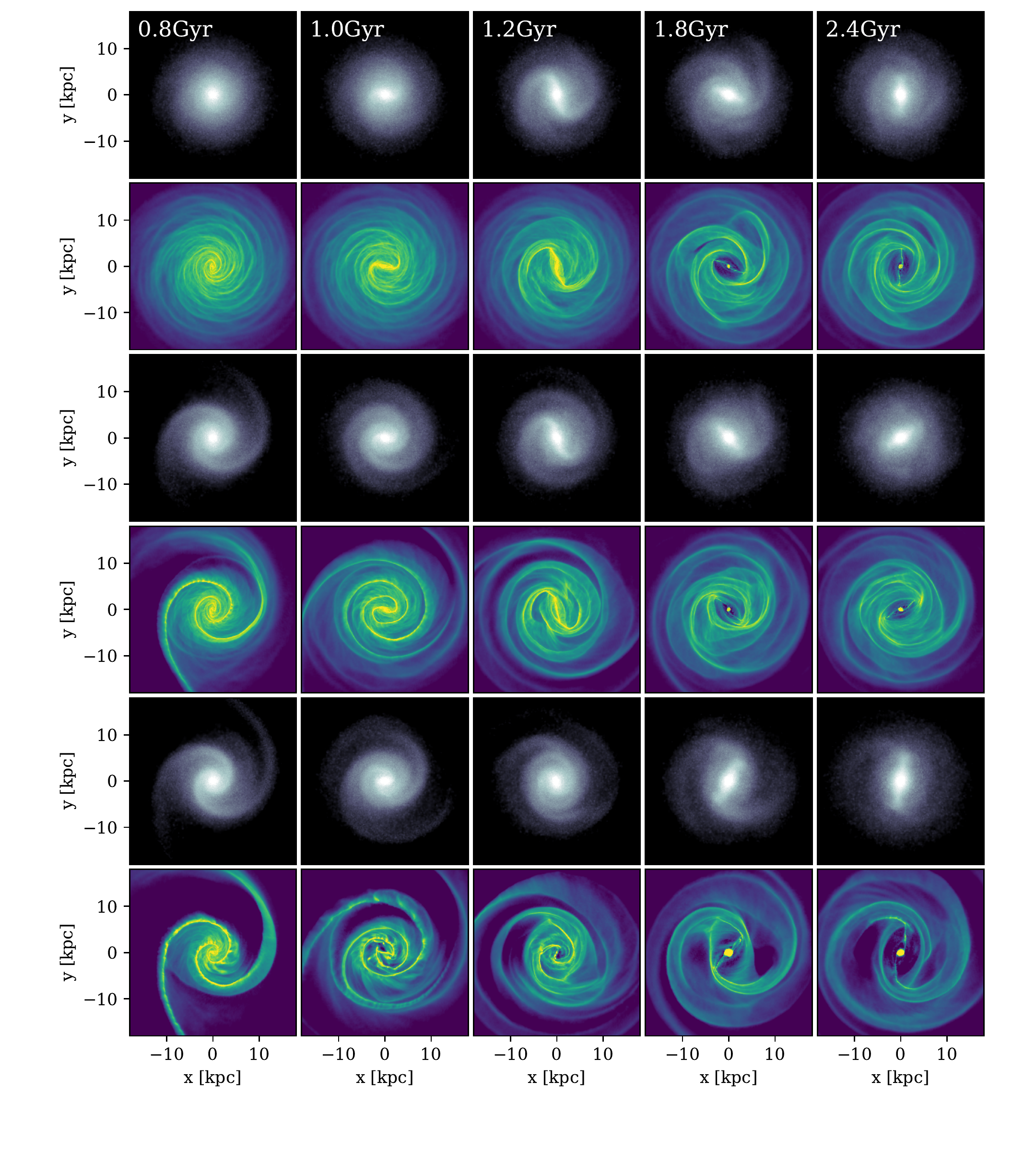}}
\caption{As Fig. \ref{IsoGalsA} but for the Mid calculations.}
\label{IsoGalsD}
\end{figure*}

\begin{figure*}
\centering
\resizebox{.90\hsize}{!}{\includegraphics[trim = 0mm 0mm 0mm 0mm]{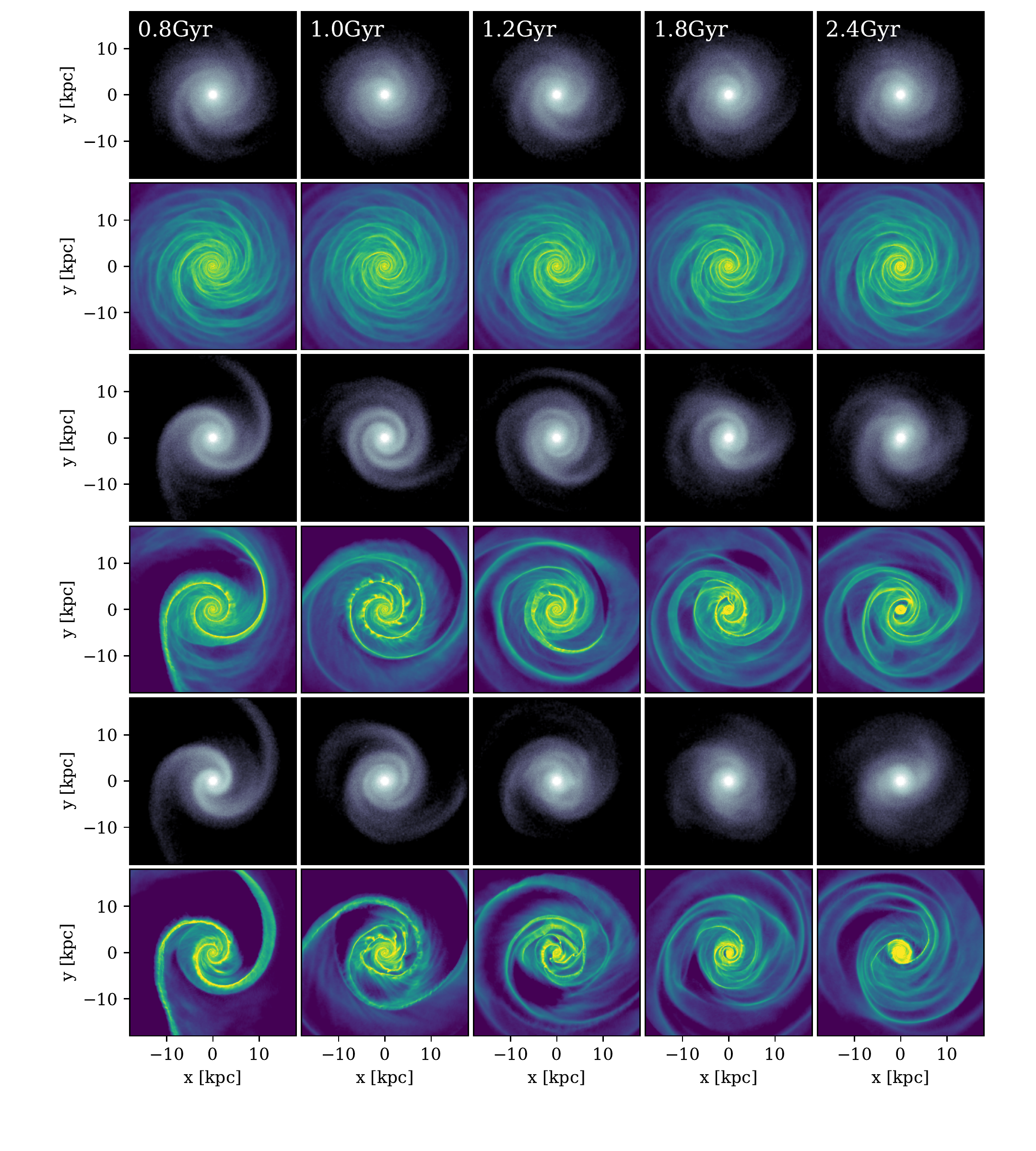}}
\caption{As Fig. \ref{IsoGalsA} but for the Dip calculations.}
\label{IsoGalsE}
\end{figure*}

Face-down views of the five isolated galaxy models are shown in the top two rows of Figures\;\ref{IsoGalsA} to \ref{IsoGalsE} in both stellar (first row) and gaseous (second row) components\footnote{Videos of all simulations in this paper can be found in the YouTube playlist: \url{https://www.youtube.com/playlist?list=PLQKy--XcWrIVBc1sS2RNc-ekyfeBsGtDs}}. Timestamps are shown that highlight changes in morphology, with the first three showing the time period corresponding to immediately after the interaction period for the S10 and S05 models, and the last two showing the longer term evolution. We briefly describe the key features below, leaving a detailed analysis of the spiral and bar features to later sections.

Each disc shows clear defining features. The most striking are the inner bars, which are readily formed in the Fall, Rise and Mid models. This is a combination of the Fall and Mid models having relatively high disc to halo mass ratios, and Rise lacking a significant inner bulge (and thus limits the existence of an ILR to damp/absorb wave propagation inwards; \citealt{1981seng.proc..111T}) to impede bar formation. The Flat and Dip models tend to form more multi-armed discs due to the lack of an inner bar driving $m=2$ outer arms, with Dip showing slightly stronger arm features.

The Flat model (Fig.\;\ref{IsoGalsA}) displays a multi-armed structure in isolation, and a clear two-armed spiral that extends to the inner disc in the interacting cases. These spirals appear to dominate the disc structure for 2-3 disc rotations before being wound up. Both interaction models show what appears to be the beginning of a bar being formed in the centre at late times, with the S10 bar appearing larger with radially extended $x_2$ orbital structures.

The Fall model (Fig.\;\ref{IsoGalsB}) has a the most centrally concentrated baryon distribution of the five models. A 2--3 armed structure is evident in the gas and stars in the isolated case until a strong inner bar is formed approximately halfway through the simulation. The interactions induce a two-armed spiral, but it does not appear as strong as the other models, and winds up very rapidly due to the Keplarian-like nature of the rotation curve. Both interactions also form short bars in the centre, but with slightly different morphology (especially in the gas).

The halo-dominated Rise model shows striking differences to the other calculations (Fig.\;\ref{IsoGalsC}). The relatively light disc shows only weak spiral structure before the formation of very strong inner bar. Both the strong and weak interactions show similar bar formation, though the bar structure seems to be highly chaotic after the interaction, changing length and shape as the tidal arms wind-up.

Mid has a similar morphology to Fall, though the weaker inner bulge allows for a more rapid bar formation in all cases (Fig.\;\ref{IsoGalsD}), owing to the absence of the wave-dampening Q-barrier from a bulge that delays bar formation. The bars seem to extend to a smaller fraction of the disc radius compared to Rise and Fall, and the interacting models appearing to create a longer bars than the isolated case.

Finally, the Dip model appears similar to Flat, but with slightly stronger isolated and perturbed spiral features (Fig.\;\ref{IsoGalsE}). The interacting cases also appear to be in the early stages of bar formation. As with Flat, the appearance of the inner disc region shows the resulting bar structure to be much larger in the stronger (S10) interaction case compared to the weaker case (S05), implying stronger interactions induce a more rapid bar response.

The Fall and Mid models have both a low predicted swing amplified spiral mode ($m_{swing}\approx2.2$) and are predicted to be strongly bar unstable ($\epsilon_{\rm bar}<0.8$ for both). This seems to agree with the simulated discs, which quickly form two armed spirals followed by short bars both in isolation and with interacting companions. The Flat and Dip models should be more stable to bar formation ($\epsilon_{\rm bar}\geq0.8$ for both) and indeed have no bar features in the isolated cases, though the interactions show evidence of the early stages of bar formation, in concordance to both models still being below the bar stability limit ($\epsilon_{\rm bar}<1.1$). Dip however has a lower predicted swing amplified mode ($m_{\rm swing}\approx2.7$) compared to Flat ($m_{\rm swing}\approx3.4$), which while not entirely clear from the face-on maps, the Fourier decomposition does show the Flat model having more power in the higher modes (e.g. $A_5$ in Fig.\,\ref{A1plt} in the following section). Rise seems somewhat of an oddity, as it should be as stable to bar formation as Flat ($\epsilon_{\rm bar}=0.9$ for both), but forms a bar on the same time-scale as Fall (which has $\epsilon_{\rm bar}=0.8$). This agrees with other studies in the literature that find $\epsilon_{\rm bar}$ and other simple bar-stability criterion may not be as suitable for all types of bars \citep{2013MNRAS.434.1287S,2016ApJ...819...92S}. Slowly rotating long bars such as those in Rise do not conform as well to such criteria as the shorter more rapidly rotating ones, and such criteria take no account of the presence of a dissipative gas disc.
 
Some calculations show similarities to the galaxies their rotation curves are based on. The Flat model has formed no bar after 2\,Gyr, similar to the modelled galaxy rotation curve (M31), but does not show the tight arm/ring system of M31, though it is believed that these observed features stem from a puncture-like interaction with the disc, not included here \citep{2014ApJ...788L..38D}. The Fall model seems very different to the flocculent nature of NGC\,4414 at later times, but these hydrodynamics + gravity only models are unlikely to fully encapsulate the complexity of such a gas-rich system. The Rise model is not unlike M33 at early times, though creates a large inner bar, quite dissimilar to M33, and it is likely that the complex nature of M33 is driven by a combination of stellar feedback and stellar disc instabilities (Dobbs et al. in preparation). The Mid model shows little in common with M81, however the morphology of the M81 system is believed to be influenced by the many members of the M81 group \citep{1999IAUS..186...81Y}. Finally, the Dip calculation is a multi-armed disc, not dissimilar to what we expect from the Milky Way system, and has the capacity to form bar features when perturbed (the Milky Way is believed to be a multi-armed barred-spiral; \citealt{2009PASP..121..213C,2014arXiv1406.4150P}).

\begin{figure}
\centering
\resizebox{1.0\hsize}{!}{\includegraphics[trim = 5mm 0mm 0mm 0mm]{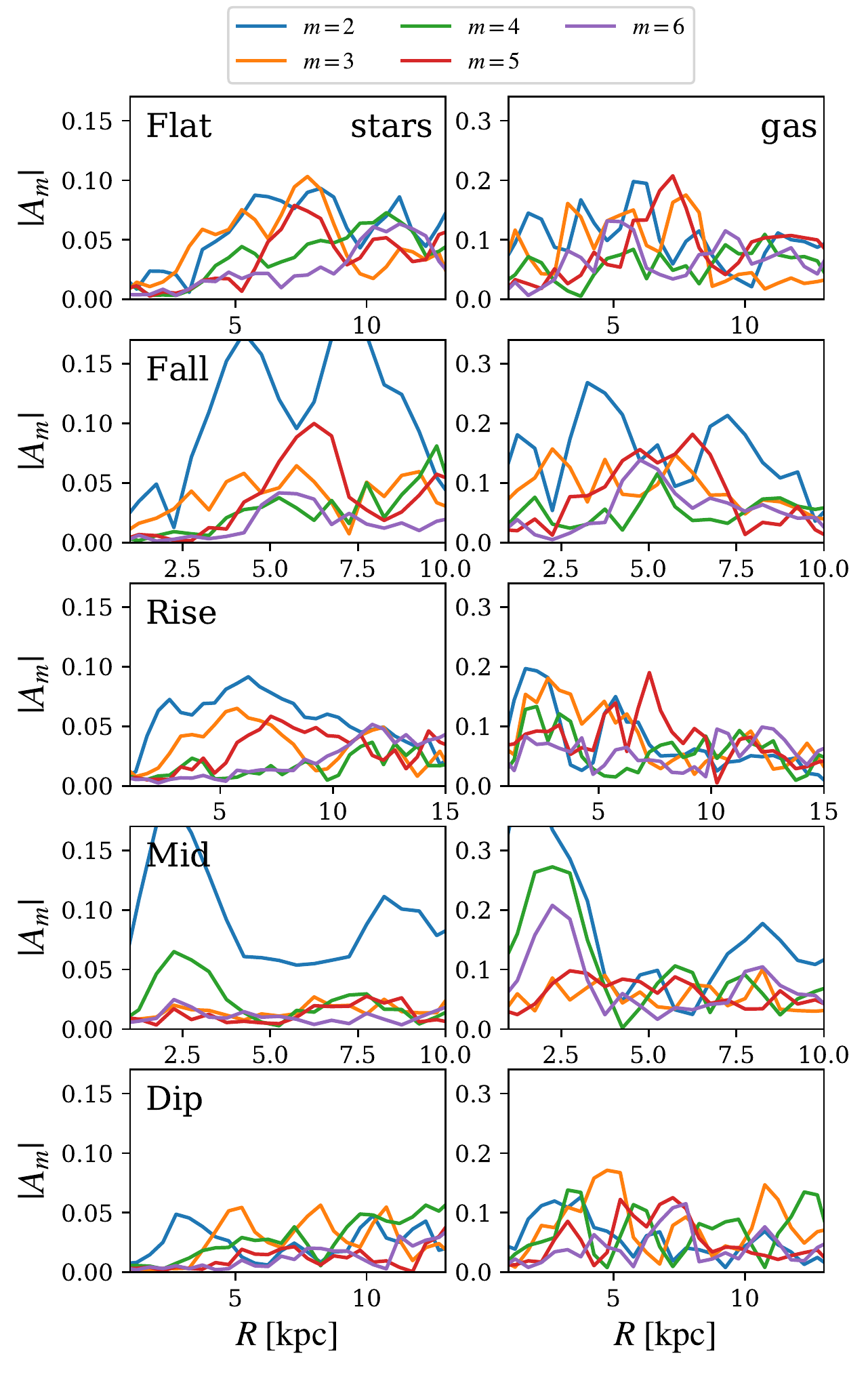}}
\caption{Magnitude of the Fourier mode amplitudes, $A_m$, of the $2\leq m\leq6$ modes of the isolated galaxies at 1.0\,Gyr, as shown in the second column of Figures \ref{IsoGalsA} to \ref{IsoGalsE}. The vertical scale for the gas is double that of the stars due to more filamentary-like gas features in comparison.}
\label{FourierIso}
\end{figure}

%%%%%%%%%%%%%%%%%%%%%%%%%%%%%%%%%%%%%%%%%%%%%%%%
\subsubsection{Mode analysis}
%%%%%%%%%%%%%%%%%%%%%%%%%%%%%%%%%%%%%%%%%%%%%%%%

In order to quantify the structures of each model, we compute a Fourier decomposition of the material in each disc, as in \citealt{2015MNRAS.449.3911P}. The Fourier $a_m$ and $b_m$ coefficients as a function of galactocentric radius, $R$, and time, $t$, for each order symmetry, $m$, are given by:
\begin{equation}
a_m(R,t)=\sum_i^{N_{\rm bin}} \cos (m\theta_i)
\label{F_am}
\end{equation}
\begin{equation}
b_m(R,t)=\sum_i^{N_{\rm bin}} \sin (m\theta_i)
\label{F_bm}
\end{equation}
where particles are binned into $N_{\rm bin}$ bins equally spaced in $R$ and $\theta$. Furthermore, the amplitude of a given mode can be calculated form:
\begin{equation}
A_m(R,t) = \frac{1}{a_0(R,t)}\sqrt{a_m(R,t)^2+b_m(R,t)^2}
\end{equation}
which is a useful parameter for characterising the strength of a specific mode as a function of radius (e.g. $A_2$ is an indicator of the bar strength in the inner disc).

In Figure \ref{FourierIso} we show the $A_m$ parameter as a function of radius for $2\leq m\leq 6$ for the five different mass models in gas and stars for the isolated cases after 1\,Gyr of evolution. We do not show such plots for the interacting cases as the $A_2$ parameter simply dwarfs the other modes at most times. In general the gas response traces features similar to those seen in the stellar material, and the dominant value of $A_m$ increases with increasing radius. All discs have the majority of their power in the $m=2$ and $3$ modes, which agrees with the predicted $m_{\rm swing}$ in Table\,\ref{Models}. Flat and Dip have no clear dominating mode, with considerable power in $m=2-4$, which is similar to other studies of dynamic spiral arms in isolated galaxies \citep{1984ApJ...282...61S,2011ApJ...730..109F,2013A&A...553A..77G}. The Fall disc has a large amount of power in $A_2$, but appears bimodal. The inner disc region is likely betraying the emergence of a bar which appears a few rotations later, while the outer arms are disconnected dynamic spirals (the disc has a predicted $m_{\rm swing}\approx2.2$). Rise also has an inner dominance of the $m=2$ mode, again signalling the coming of a bar only a rotation after, while the outer disc appears effectively flocculent. At 1\,Gyr, the Mid model has already formed a bar, and so the entirety of the spectrum is dominated by the $m=2$ mode.

\begin{figure}
\centering
\resizebox{1.0\hsize}{!}{\includegraphics[trim = 7mm 0mm 2mm 0mm]{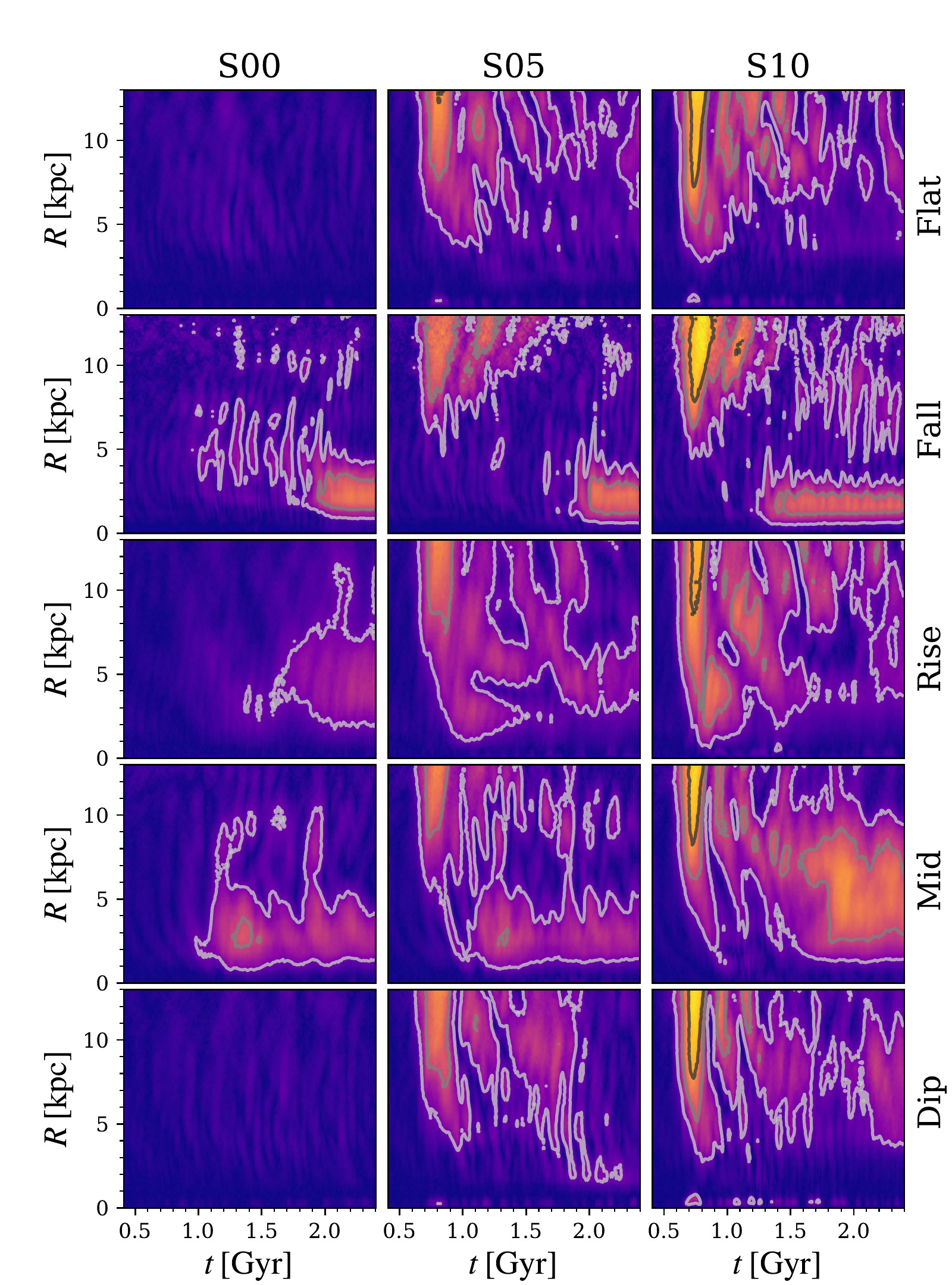}}
\caption{Map of the $A_2$ mode power as a function of time and radius for all 15 models. Each row shows a different mass model, and each column a different interaction scenario. The colours render shows mode power in the range of $0.0\leq A_2 \leq 0.8$. Grey-scale contours are drawn at powers of 0.2, 0.4 and 0.6.}
\label{A2map}
\end{figure}

\begin{figure}
\centering
\resizebox{1.0\hsize}{!}{\includegraphics[trim = 0mm 0mm 0mm 0mm]{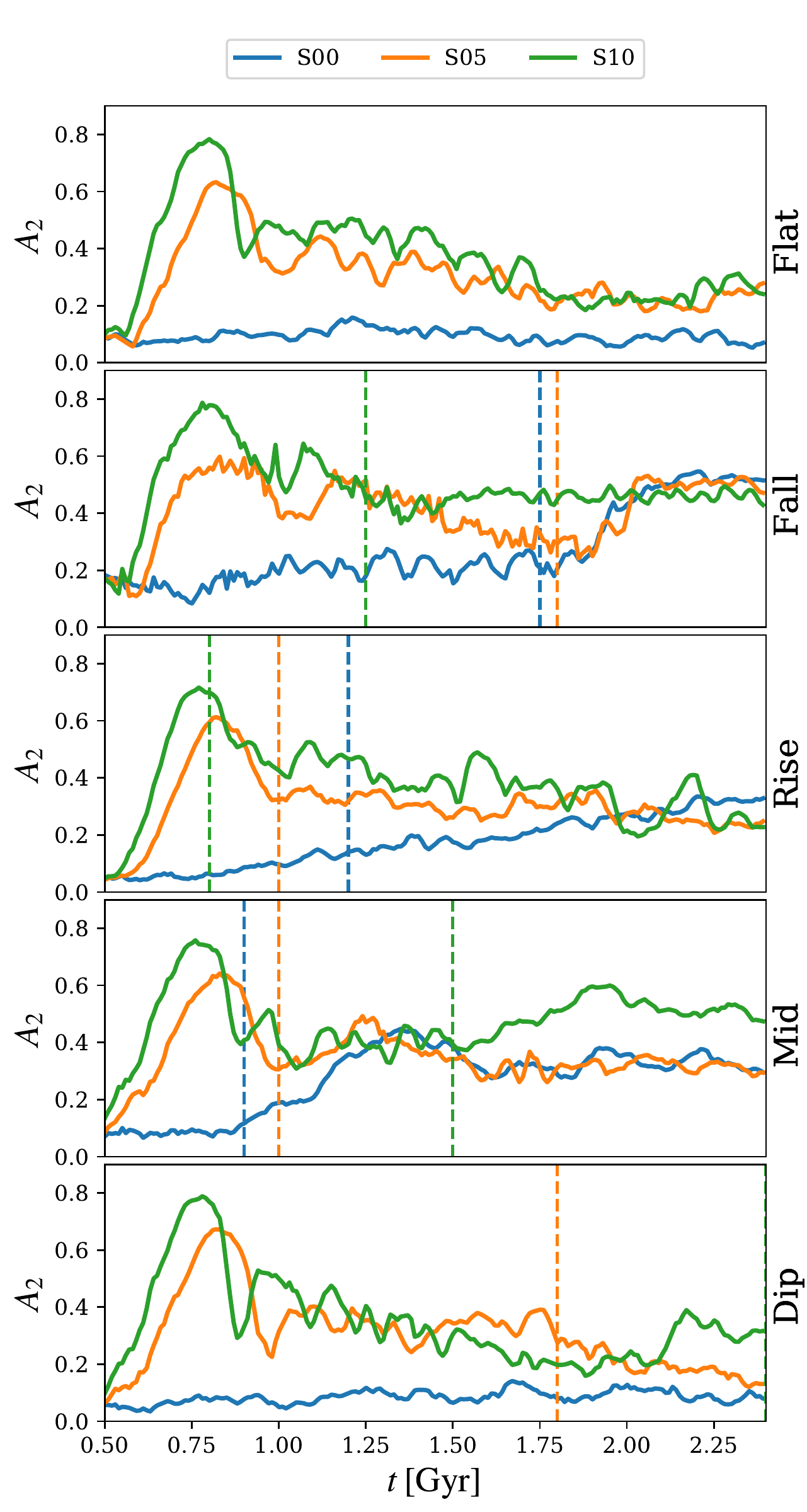}}
\caption{Power of the $A_2$ mode power as a function of time for all 15 models. Each panel shows a different mass model, with different coloured lines indicating a different interaction scenario. The vertical dashed lines show the approximate bar formation times for each of the models where a bar is present (coloured coded the same as the solid lines).}
\label{A1plt}
\end{figure}

In Figure\;\ref{A2map} we show $A_2$ in the stars as a function of radius and time for all calculations, with values shown in the range of $0.\leq A_2 \leq 0.8$. Each row denotes a different mass model, and each column a different interaction scenario. In addition, the maximum power of $A_2$ as a function of time for all models is shown in Figure\;\ref{A1plt}, with each colour representing a different interaction scenario. Vertical lines indicate the approximate bar formation timescale (see \ref{sec_bar} for our definition of bar structure). Both figures \ref{A2map} and \ref{A1plt} show a sharp increase in $A_2$ in the mid--outer disc after closest approach in the interaction models, indicating the formation of the $m=2$ tidal spiral arms. In the Flat model there is no clear change in $A_2$ for S00, but the companion runs show a clear increase and then gradual drop-off in $A_2$ after the companion closest approach (Fig.\;\ref{A1plt}). Interestingly the power does not drop back to pre-interaction levels and instead plateaus at around $A_2=0.3$. This is contrary to the results of \citet{2016MNRAS.458.3990P}, where the $A_2$ power dropped back to pre-interaction levels after 1\,Gyr. This directly shows the impact of having a disc that is more bar unstable, as the model from \citet{2016MNRAS.458.3990P} had a large inner bulge and centrally concentrated halo designed to heavily impede bar growth, so the disc was able to settle back to its original morphology after the fly-by.

The Fall model shows S00 rising slowly over time early in the simulation. As the bar is formed around 1.8\,Gyr $A_2$ rapidly rises, plateauing near the end of the simulation. At this point the disc has a clear bar and well defined resonance regions (i.e. outer and inner rings). The weak interaction (S05) shows $A_2$ rise, slowly drop after the interaction, and then abruptly rise again once the bar is formed in a similar manner to the isolated case. This implies the weak interaction has no impact on the bar strength or formation time for this model. Figure\;\ref{A2map} clearly shows the bar in the bottom right corner of S05 and S00, and it being of roughly equal radial extent and formation epoch, disconnected form the outer arm features. The S10 model forms the bar much earlier, though the final power of $A_2$ is the same level as the S05 and S00 cases. In this case the strong interaction has only accelerated the bar formation, with the final strength being the same across all interaction scenarios (Fig.\;\ref{A1plt}).

The RiseS00 model shows a slow and steady increase in $A_2$ as the bar is slowly formed. The interaction cases show $A_2$ slowly dropping after the interaction, and plateauing at the same value at the end of simulation, much like the Flat S05/S10 models even though RiseS05/S10 have formed a strong inner bar. The maps of $A_2$ (Fig.\;\ref{A2map}) show the chaotic nature of the bar in RiseS05/10, with no clear ridge in power as seen in Fall. Constant disconnects between outer and inner material make the bar a more fluctuating feature than the bars of Fall.

Mid shows similar behaviour as Fall, with the bar epoch characterised by a growth stage in $A_2$ shortly after formation in the isolated case. The weak interaction similarly reaches the same power once the bar has been formed. The strong interaction, however, shows the bar grows much stronger than the S05 and S00 cases. This is clear evidence that stronger interactions generate stronger bars, though this bar takes longer to emerge than the S05/S00 cases. MidS10 shows clear connections from the outer tidal spirals to the inner bar regions in Figure\;\ref{A2map}, also visible to come extent in RiseS05, which appear to have fuelled the growth of the bar compared to MidS00 and MidS05. Such features in $A_2(t,R)$ space were also seen in \citet{2017A&A...604A..75M}.

Finally, Dip is very similar to Flat, though the proto-bar structures appear more advanced than those of Flat. This is seen in the Figure\;\ref{A1plt} data for S10, which begins to abruptly rise near the end of the simulation. Figure \ref{A2map} also shows this ridge of power emerging around 8kpc in DipS01 in contrast to the other Dip models, indicating the emergence of a large bar structure.

These results suggest that interactions induce bar formation early and produce bars with greater strength, though not consistently between all models, with Mid being the particular anomaly. The Flat and Dip curves clearly have some early-stage bar features in the gas as a result of the companion passage, highlighted by their $A_2$ power not dropping back to the isolated value almost 2\,Gyr after the interaction, with these appearing stronger in S10 than S05. Rise and Fall show an acceleration in bar formation time, though their bars appear similar in strength to the isolated case. All models have clear tidal spiral arms that manifest over 0.25Gyr and dissipate over the course of a 1Gyr, some of which appear to directly interact with the young bar features.

%%%%%%%%%%%%%%%%%%%%%%%%%%%%%%%%%%%%%%%%%%%%%%%%
\subsection{Pattern speeds}
%%%%%%%%%%%%%%%%%%%%%%%%%%%%%%%%%%%%%%%%%%%%%%%%

There are two popular methods for defining the pattern speed of bar/spiral structure within a simulation. The first is to directly trace the peaks in density as a function of azimuth and then simply track their motion over time at any given radius (e.g. \citealt{2012MNRAS.426..167G,2015MNRAS.449.3911P}). Peaks can be extracted by a spatial Fourier analysis of the disk at a given time for a specific mode, $m$, where the particles have been binned into $R-\theta$ space (using $a_m$ and $b_m$ from equations\,\ref{F_am} and \ref{F_bm}). The change in $\theta$ of each peak for a given $m$ then gives the pattern speed of that specific mode. The other method involves performing a power spectrum analysis in time \citep{1987MNRAS.225..653S,2011MNRAS.417..762Q,2012MNRAS.426.2089R}. The power spectrum, $P_m$, can be calculated over a range of frequencies up to the Nyquist frequency via:
\begin{equation}
P_m(R,\omega)=\int_{t_1}^{t_2}[a_m(R,t)+ib_m(R,t)]h(t)e^{i\omega t}dt
\label{spec}
\end{equation}
where $h(t)$ is some window function to reduce aliasing (e.g. Gaussian, Hanning). Spectrograms can then be plotted of $|P_m(R,\Omega_p)|$, where the pattern speed is given by $\Omega_p=\omega/m$. Ridges in these spectrograms should indicate the pattern speeds of each mode in a given time frame.

\citet{2013A&A...553A..77G} discuss the merits of both of these methods in the context of dynamic spiral arms in an isolated galactic disc. They find that while the direct method gives a much better time resolution and thus is better for more transient features, the mode pattern analysis allows for a clearer decomposition of underlying pattens in the disc that may occupy the same position at any snapshot (see also \citealt{2013MNRAS.432.2878R}).

\begin{figure}
\centering
\resizebox{1.0\hsize}{!}{\includegraphics[trim = 5mm 5mm 5mm 0mm]{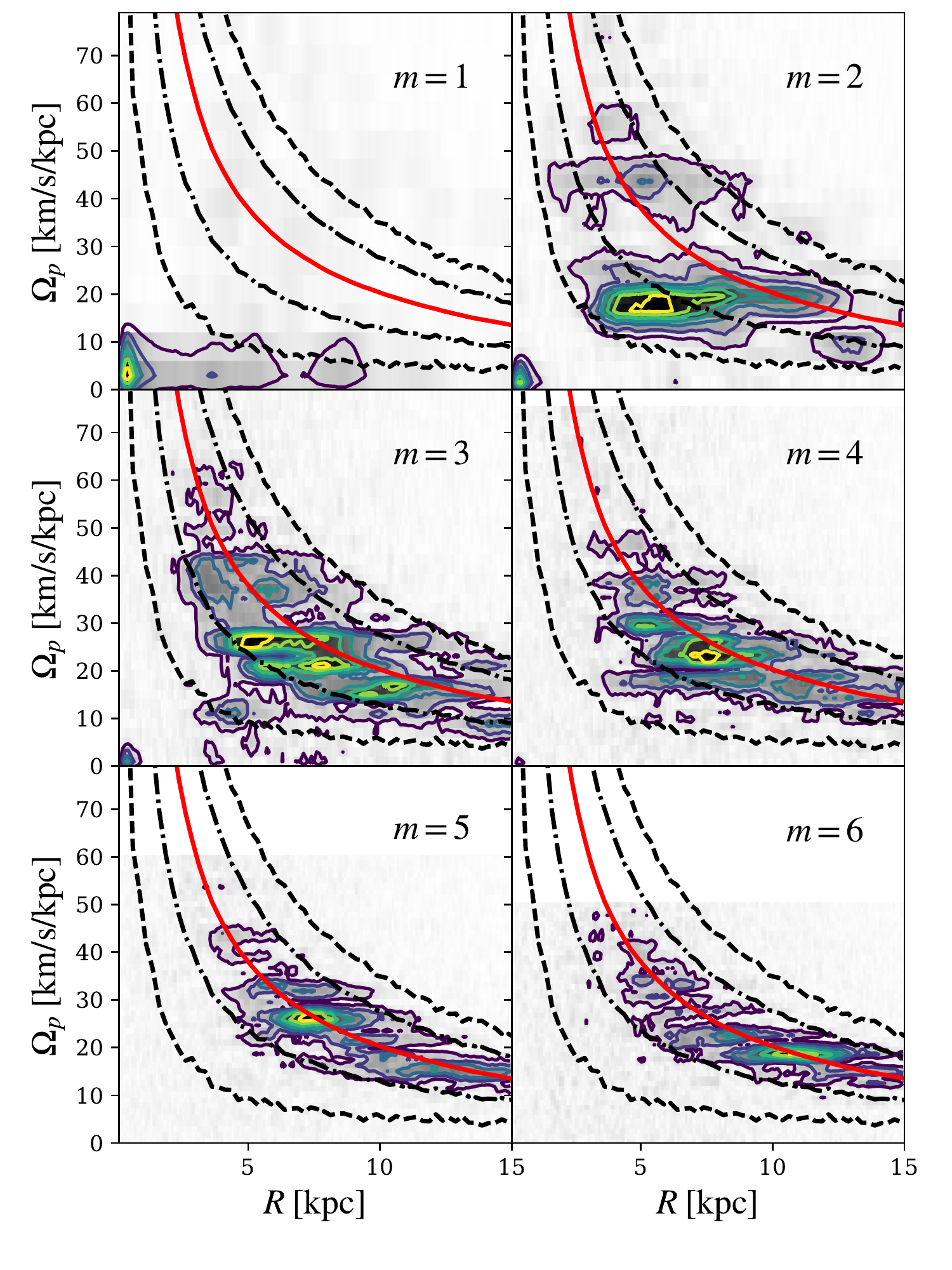}}
\caption{Spectrograms of the stellar power defined by Eq.\;\ref{spec} over a range of frequencies and radii in the FlatS00 calculation. Each panel represents a different mode, $m$. The dashed, dashed-dot and red solid lines indicate the $\Omega\pm\kappa/2$, $\Omega\pm\kappa/4$ and co-rotation frequencies inherent to the disc. The Nyquist frequency determines the upper limit of frequencies visible in the $m=5$ and $m=6$ panels.}
\label{spec00}
\end{figure}

As we are mostly interested in global patterns in the discs, rather than tracing specific spiral arm features (which was more the focus of \citealt{2016MNRAS.458.3990P}) we chose the spectrogram analysis as our primary means for determining pattern speeds. An example of such spectrogram analysis is shown for the FlatS00 calculation in Figure\,\ref{spec00}, with dashed lines indicating the $\Omega\pm\kappa/2$ resonances, dashed-dot lines showing $\Omega\pm\kappa/4$, and red solid lines showing co-rotation. We calculate the power spanning 1\,Gyr of the disc in isolation, sampling the simulation every 10\,Myr. Each panel represents a different mode, in the range $1\leq m\leq 6$. This calculation shows clear features that appear to rotate as material waves for all $m\geq 3$ modes, with a pattern speed that rotates seemingly equal to or just less than the co-rotation frequency. There is also a clear trend in higher mode numbers dominating the outer regions of the disc, as predicted by Eq.\,\ref{SwingAmp} (see also the $m=2$ and $m=9$ spectrograms of the interactions of \citealt{2016MNRAS.461.2789H}). The $m=2$ mode shows a clear horizontal ridge in frequency space, highlighting the underlying bar-forming potential of the disc, though none is clearly manifest in the disc at this time in the face-on plots of Fig.\ref{IsoGalsA}.

\begin{figure}
\centering
\resizebox{1.0\hsize}{!}{\includegraphics[trim = 5mm 10mm 5mm 0mm]{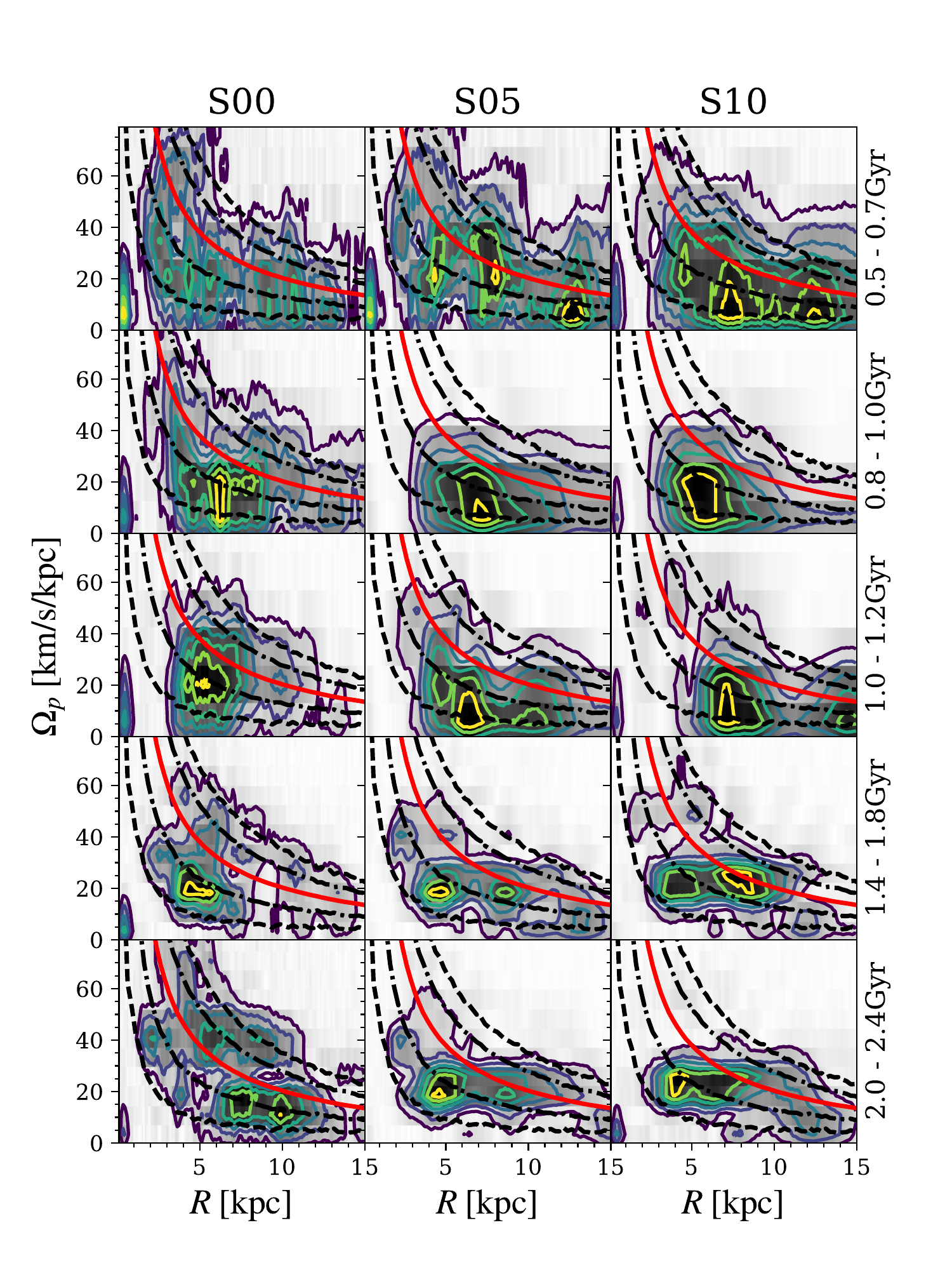}}
\caption{Spectrograms of the stellar power in the $m=2$ mode in all models using the Flat rotation curve. The left, centre and right columns show the isolated, weak interaction, and strong interaction cases. Five different time ranges are shown, increasing from top to bottom, with the 0.8--1.0\,Gyr row showing the peak interaction response.}
\label{spec1}
\end{figure}

\begin{figure}
\centering
\resizebox{1.0\hsize}{!}{\includegraphics[trim = 5mm 10mm 5mm 0mm]{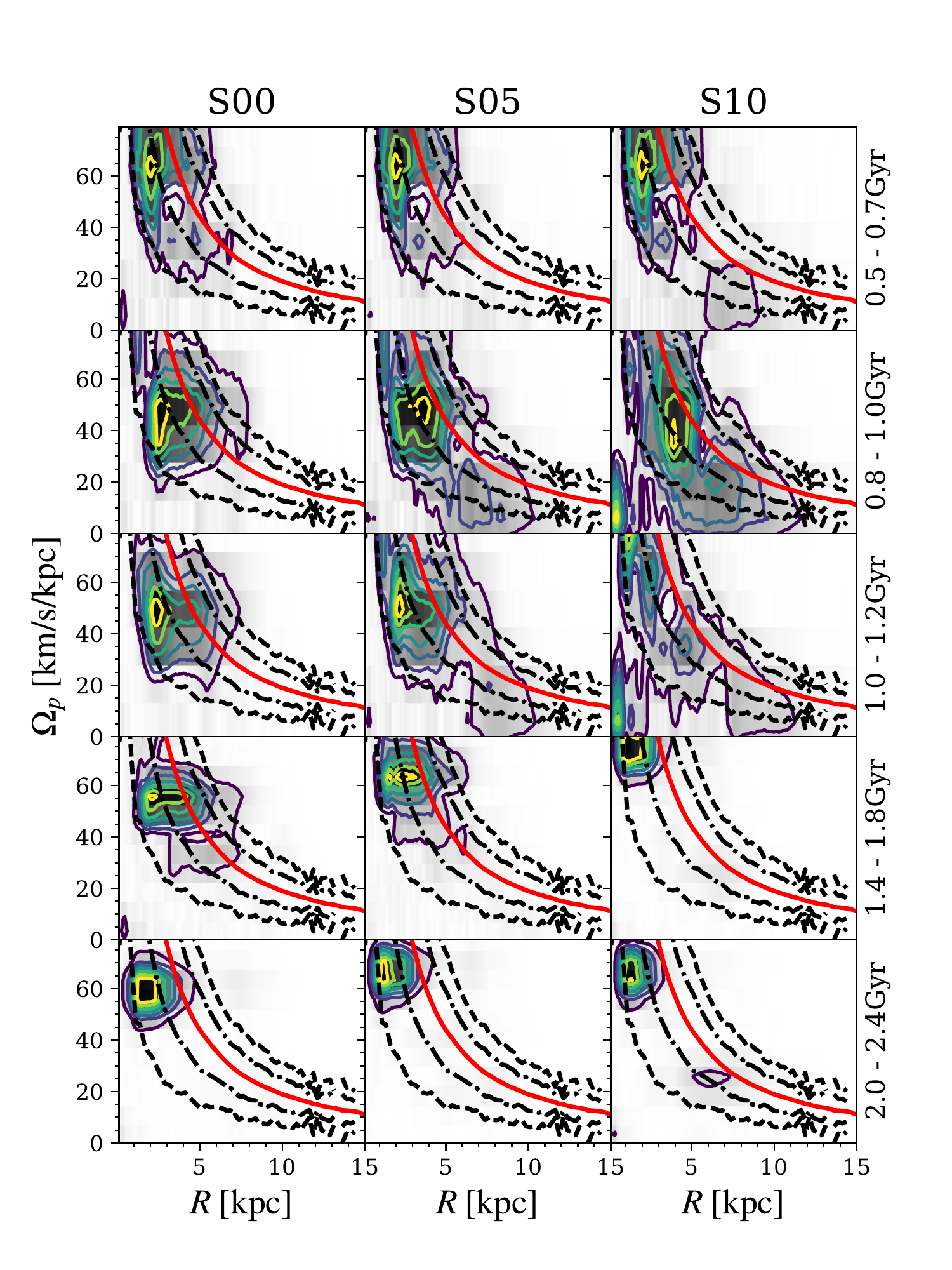}}
\caption{As Figure \ref{spec1} but for the Fall calculations.}
\label{spec2}
\end{figure}

\begin{figure}
\centering
\resizebox{1.0\hsize}{!}{\includegraphics[trim = 5mm 10mm 5mm 0mm]{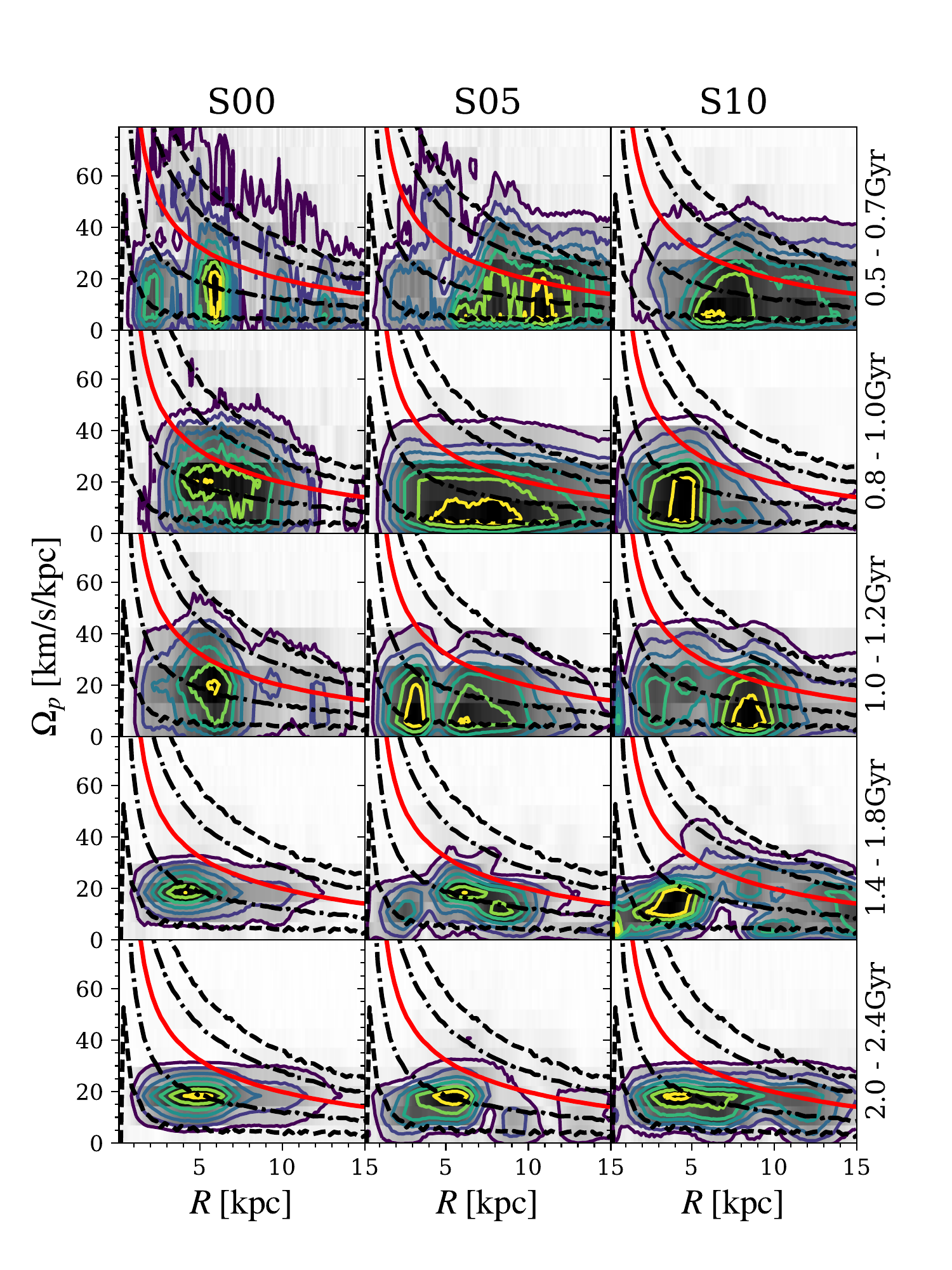}}
\caption{As Figure \ref{spec1} but for the Rise calculations.}
\label{spec3}
\end{figure}

\begin{figure}
\centering
\resizebox{1.0\hsize}{!}{\includegraphics[trim = 5mm 10mm 5mm 0mm]{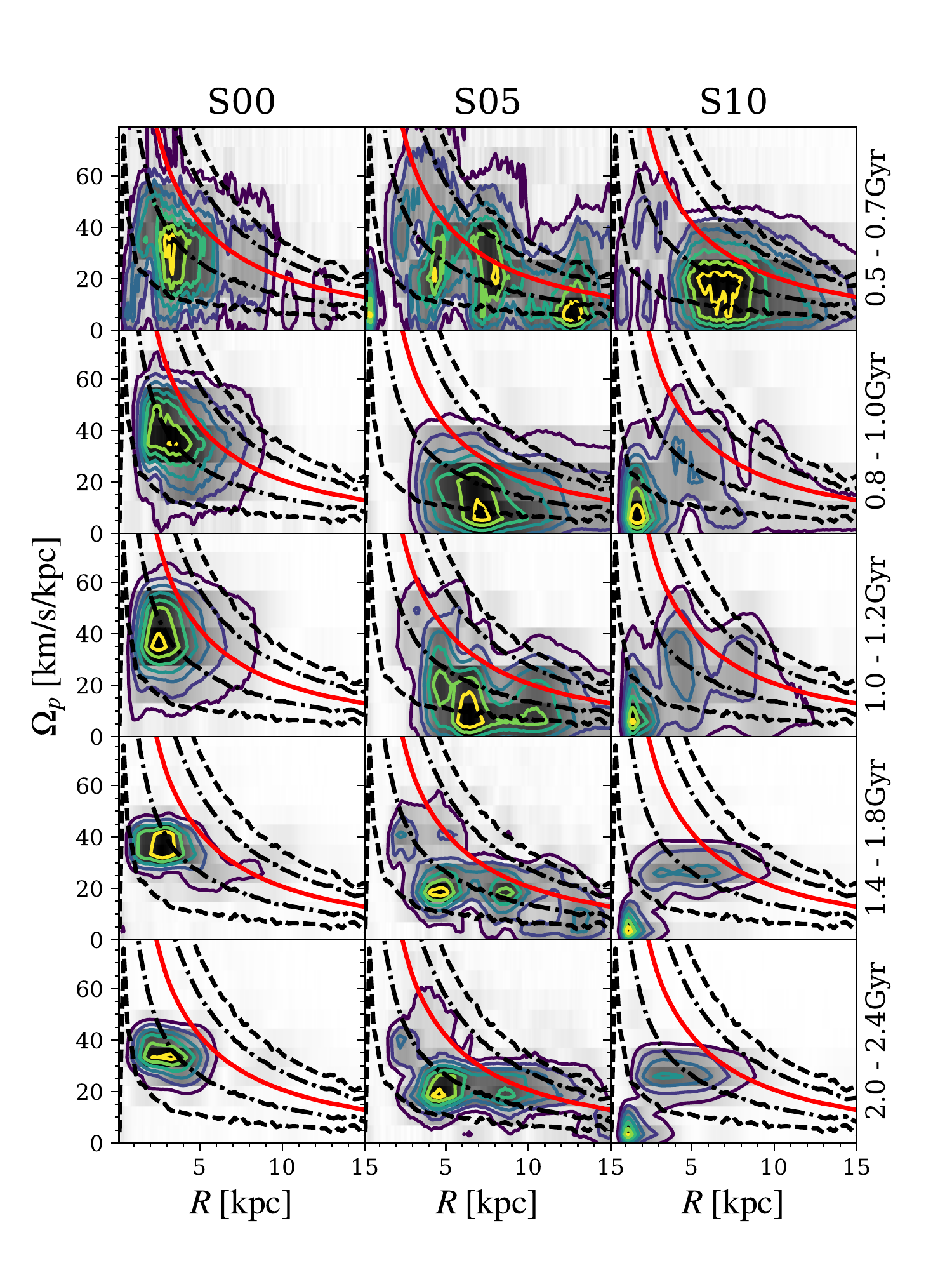}}
\caption{As Figure \ref{spec1} but for the Mid calculations.}
\label{spec4}
\end{figure}

\begin{figure}
\centering
\resizebox{1.0\hsize}{!}{\includegraphics[trim = 5mm 10mm 5mm 0mm]{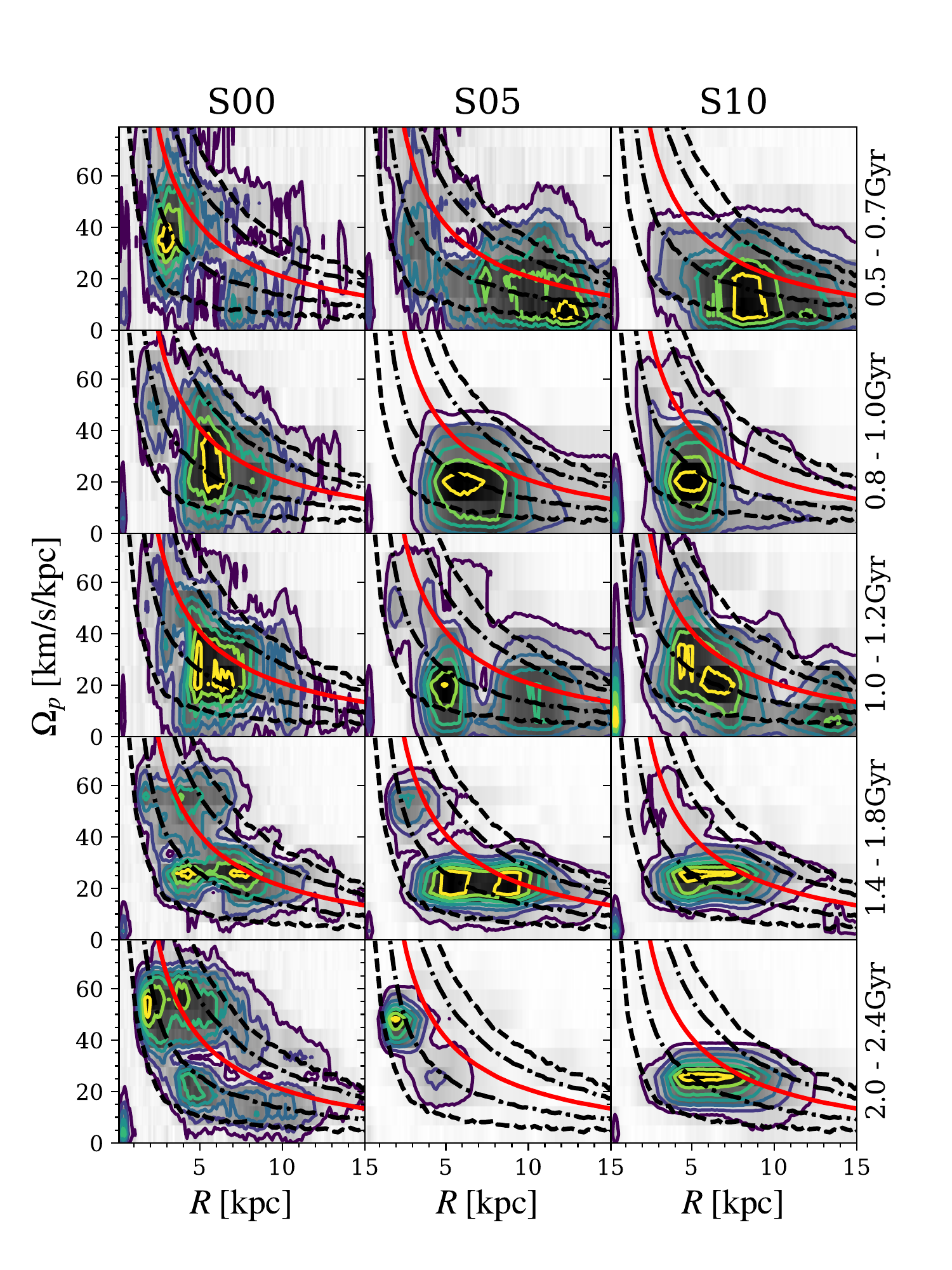}}
\caption{As Figure \ref{spec1} but for the Dip calculations.}
\label{spec5}
\end{figure}

For the remaining models we will only show relevant modes that warrant discussion, namely the $m=2$ mode, which traces both tidal spirals and bars. Figures\,\ref{spec1}--\ref{spec5} show the spectrograms of each mass model, with each column showing a different interaction scenario and each row a different epoch. The red and black lines have the same meaning as in Figure\,\ref{spec00}, though are obviously different in each figure as they are a direct product of the shape of the rotation curves.

Figure\;\ref{spec1} shows spectrograms of the $m=2$ mode in the Flat model. The isolated model appears to have a disperse $m=2$ feature at all times, though the exact structure of the feature seems to move around in $\Omega_p$--$R$ space over the course of the simulation. The pattern does not seem strongly correlated to co-rotation (red line) and the transient nature is likely because the $m=2$ pattern is not the dominant feature in the disc (see Fig.\,\ref{FourierIso}), favouring instead the $m=3$ mode. For the S05 and S10 interactions there exists a strong $m=2$ feature, especially clear just after perigalacticon (0.8--1Gyr). The pattern speeds lie around the Inner Lindblad Resonance (ILR)/2:1 regime, and move up to the 4:1 frequencies over time. Towards the end of the simulation, the $m=2$ mode shifts up into a horizontal ridge, signalling the precursor of some steady underlying density wave. Face-on maps indicate this power is in fact associated with a spiral rather than bar at this time (see also Sec.\;\ref{sec_bar}) though this phase is in likelihood not stable and will soon give way to a barred structure \citep{2011MNRAS.410.1637S,2013MNRAS.432.2878R}. Differences between S10 and S05 are minimal, with S10 having a slightly cleaner $m=2$ signal. The interesting result of this model is that the companion has induced some two-armed spiral feature long after the initial tidal spiral feature has wound up, and that this feature is not evident in the isolated disc nor is it a clear bar at this epoch.

Spectrograms for the baryon dominated Fall models are shown in Figure\;\ref{spec2}. These models show a much clearer evolution. All three appear to develop a strong inner $m=2$ feature early, indicating the presence of a short, rapidly rotating inner bar progenitor. For the S05 and S10 calculations, the companion drives an increase in $m=2$ power in the mid disc, though this rapidly dissipates to leave the spectrogram relatively unchanged compared to the pre-interaction state. The main difference is that the peak power is slightly shifted vertically upwards to higher frequencies in the interacting cases, implying the bar feature has gained some angular momentum from the companion.

The dark matter dominated Rise model pattern speeds are shown in Figure\;\ref{spec3}. The interactions clearly induce a very slow rotating outer $m=2$ pattern aligned with the 2:1 resonance that is associated with the tidal spiral arms. The S00 model has a slower build up of $m=2$ power that initially rotates with the 4:1 resonance until it levels out to a constant pattern speed of approximately 20\ps{} by the end of the simulation. This power stems from a large and slowly rotating inner bar, which converges to the same pattern speed in all cases, rising up from the 2:1 frequencies in the interacting galaxies and dropping down form the 4:1 frequencies in the isolated galaxy.

Figure\;\ref{spec4} shows the Mid calculations, characterised by their disc dominated mass model. The isolated model shows an $m=2$ pattern that converges from the 4:1 resonance into a constant pattern speed feature. At intermediate times in the early life of the bar, there are strong spiral arms emanating from the bar ends, creating the bleeding out of $m=2$ power into a large region of $\Omega_p$--$R$ parameter space. This dissipates at later times, leaving only the bar. The weak interaction has a smooth transition from a component that traces the 2:1 resonance into an inner bar with a constant pattern speed. Interestingly the outer regions with constant pattern speed do not seem associated with the bar ($4{\rm kpc}\leq R\leq13{\rm kpc}$), which is instead the power in the region $R<5$kpc and $\Omega_p=35$\ps{}, the same as in the isolated case. This outer component stems instead from a spiral pattern that periodically connects and reconnects to the bar. This arm-bar disconnect has been observed in other simulations \citep{1988MNRAS.231P..25S,2015MNRAS.454.2954B}, and could explain why some observed galaxies have clear disconnect region between their inner bar and outer arms. The strongest interaction also has a two-armed spiral that transforms into an inner bar, though this bar is longer and rotates slower than its S00 and S05 brethren. There is also considerable power in some small feature in the lower left of the plots. This is due to a few clumps of material flowing into the galactic centre, transporting angular momentum into the bulge region, which causes the centre of the stellar disc to precess slightly around the centre of mass of the galaxy.

The final set of models, with the dipped rotation curves, are shown in Figure\;\ref{spec5}. The S00 disc has an $m=2$ feature that rotates somewhat slower than the corotation frequency, and over time levels out into a flat ridge of constant pattern speed. Like the Mid and Flat models, this produces a slow moving outer feature ($\approx 20$\ps{}) and a rapidly rotating inner proto-bar feature ($\approx 55$\ps{}) at late times. Note that these pattern speeds are in excellent agreement with the bar and spiral pattern speeds observed in the Milky Way, whose rotation curve this model is based on \citep{2011MSAIS..18..185G}. The power in the S05 and S10 interactions again appear to trace the ILR shortly after closest approach, though the end points seem very different. The weaker interaction creates a rapidly rotating short inner bar with some slower outer spirals, while the strong interaction creates a single large component with a constant pattern speed, rotating much slower than the S05 bar. In S10 the bar is still in its early stages, and appears to rotate with connected spiral arms that extend up to the disc edge. It first displays an outer tidal spiral that wraps up, giving way to a new $m=2$ component that dominates the entire disc. This feature seems moderately long lived, lasting for almost a full Gyr. The existence of such long lived spiral features is still an open question. While they appear to be reproducible to some degree in simulations, they are either precursors to bars (as is the case here) transient structures that alternate from $m=2$ to $m=3$ or superpositsions of multiple underlying waves that grow and decay independently \citep{2011MNRAS.410.1637S,2014ApJ...785..137S,2014PASA...31...35D}.

%%%%%%%%%%%%%%%%%%%%%%%%%%%%%%%%%%%%%%%%%%%%%%%%
%%%%%%%%%%%%%%%%%%%%%%%%%%%%%%%%%%%%%%%%%%%%%%%%
\subsection{Spiral structure}
%%%%%%%%%%%%%%%%%%%%%%%%%%%%%%%%%%%%%%%%%%%%%%%%
%%%%%%%%%%%%%%%%%%%%%%%%%%%%%%%%%%%%%%%%%%%%%%%%

%%%%%%%%%%%%%%%%%%%%%%%%%%%%%%%%%%%%%%%%%%%%%%%%
\subsubsection{Nature of the spirals}
\begin{figure}
\centering
\resizebox{1.0\hsize}{!}{\includegraphics[trim = 0mm 0mm 0mm 0mm]{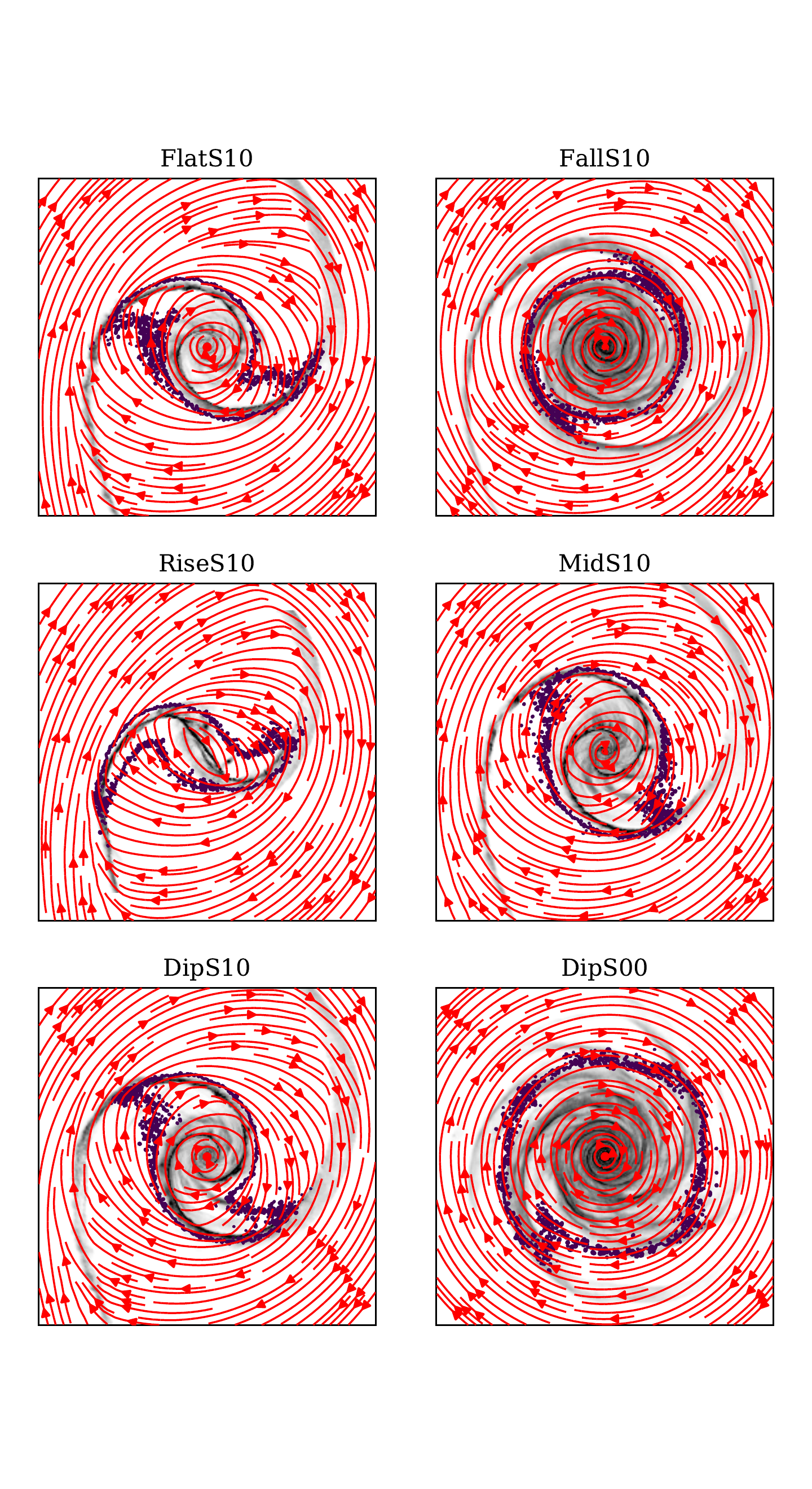}}
\caption{Maps of the stellar velocity field (red arrows) with gas density render and stellar density contour overplotted. Orbit crowding can clearly be seen along the spiral arms, especially for the barred RiseS10 model.}
\label{Orbits}
\end{figure}

%%%%%%%%%%%%%%%%%%%%%%%%%%%%%%%%%%%%%%%%%%%%%%%%
Interactions inducing $m=2$ spirals in disc galaxies have been well documented in the literature \citep{1972ApJ...178..623T,1991A&A...244...52E,1992AJ....103.1089B}. These spiral arms behave as density wave-like structures, with pattern speeds slower than the material rotation speed of the discs. However, this rotation speed is not constant, and the waves will wind-up over time, implying that tidal spirals lie between classical steady density waves and dynamic material spiral arms. Figure \ref{Orbits} shows the tidal spirals at their peak strength in each of the S10 interaction models (and the DipS00 model in the bottom right). Gas is shown in grey-scale, stellar material as a single contour, and velocity field lines as the red arrows. Clear orbital crowding can be seen in the spiral arm regions, indicating shocked regions of gas. The Rise model in particular has field lines changing orientation by nearly $90^\circ$ as they enter the spiral, which continues well into the centre where the newly formed bar has created highly elliptical orbital features. Conversely, the spiral arms in DipS00 model (which have not been created by a tidal interaction) show very weak or almost non-existent orbital crowding. The velocity field lines appear circular, regardless of the moderate strength spirals present, further highlighting differences in the nature of these spiral arms formed in isolation \citep{2010MNRAS.409..396D,2015PASJ...67L...4B}.

Evidence from past studies suggest that tidal spirals rotate at some speed between the corotation frequency and some constant speed slightly slower than the $\Omega(R)-\kappa(R)/2$ frequencies, though there is increasing evidence that these spirals rotate exactly with $\Omega(R)-\kappa(R)/2.$ \citep{2015ApJ...807...73O,2016MNRAS.458.3990P,2017ApJ...834....7S}. As the frequencies depend on the mass model and thus the rotation curve of a given galaxy, this makes our simulation suite an excellent test bed for further validating or discrediting this theory.

The pitch angle, $\alpha$, of a given spiral arm should wind-up at a rate given by:
\begin{equation}
g(\Omega)=\cot{\alpha} = \Big| Rt \frac{d\Omega_p(R)}{dR} \Big|,
\label{windEq}
\end{equation}
where $t$ is simply the time after some reference point, $R$ is the radius from the disc centre, and $\Omega_p(R)$ is the pattern speed of the rotating spiral. For material arms, $\Omega_p(R)=\Omega_{\rm disc}(R)$=$V_c/R$, giving a rapid wind-up rate. For density waves $\Omega_p(R)=\Omega_{p,0}={\rm const.}$, resulting in no winding up. For tidal arms authors postulate that the spirals rotate with the 2:1 frequency, i.e. $\Omega_p(R)=\Omega_{\rm disc}(R)-\kappa(R)/2$, which is slower than the material rate but still non-zero. 

\begin{figure}
\centering
\resizebox{0.84\hsize}{!}{\includegraphics[trim = 0mm 0mm 0mm 0mm]{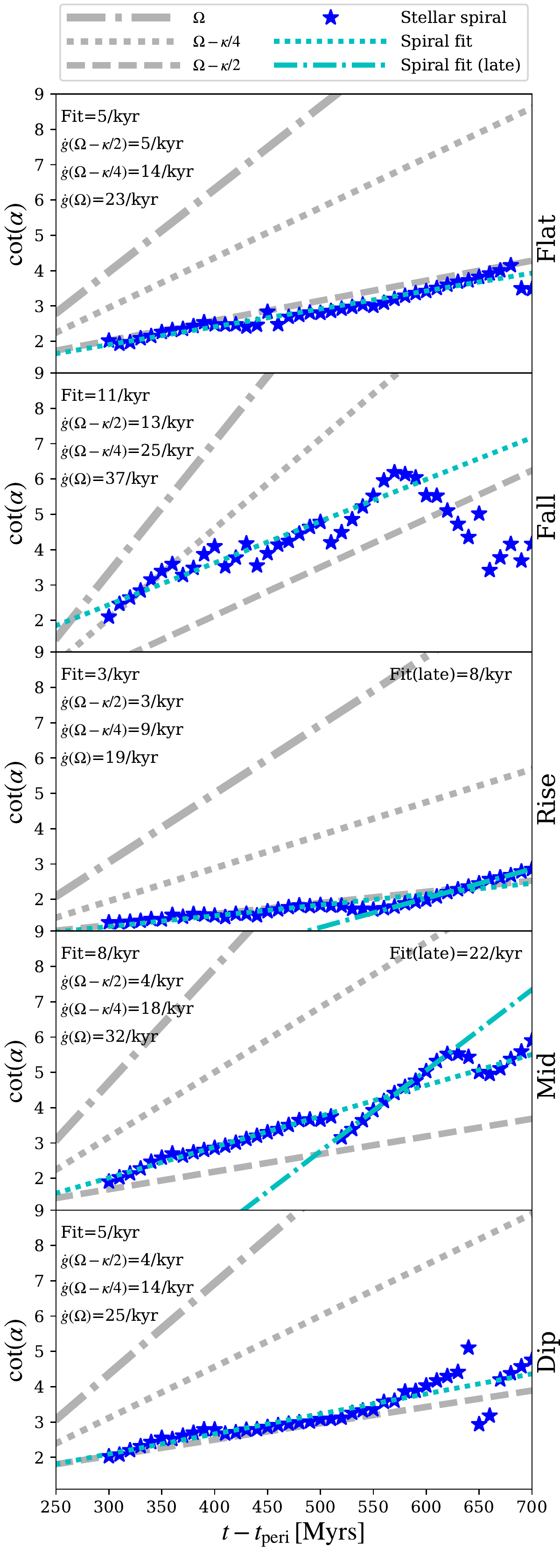}}
\caption{The wind-up rate of stellar spirals in all five S05 calculations. The cotangent of the spiral pitch angle, $\alpha$, is shown on the $y$-axis as a function of time, which increases as the spiral becomes more tightly wound. Blue stars indicated the measured pitch angles for each model. The function $g={\rm cot(\alpha)}$ of Eq. \ref{windEq} for spirals winding up at a rate of $\Omega$, $\Omega-\kappa/4$ and $\Omega-\kappa/2$ are shown as the grey dot-dashed, dotted and dashed lines. The cyan dotted line shows linear fits to the data, with the dot-dashed line showing fits at later epochs where relevant. The gradient for each wind-up model, as well as for a linear fit to the measured data, is given in the upper left of each panel.}
\label{AlphaDrop}
\end{figure}

In Figure\,\ref{AlphaDrop} we show the wind-up rate for the tidal spirals in all S05 calculations (blue started points). These spiral arms are fit using a simple direct tracing method \citep{2012MNRAS.426..167G,2015MNRAS.449.3911P}, as the temporal resolution of the spectrogram analysis in the pervious section is far too coarse to resolve such a wind-up rate. We plot $\rm cot(\alpha)$ to give a linearly increasing plot, against $t-t_{\rm peri}$, the time since perigalacticon passage of the companion. We perform linear fits to the data, shown by the cyan dotted line, with the gradient for the fit indicated in the top left of each plot. For Rise and Dip there are clear single breaks in the wind-up rate. As such we perform fits to both an early and late epoch, the fit to the late epoch is given in the top-right of the relevant panels. The Fall model displays multiple breaks, a result of the spiral perturbation struggling to penetrate the mid/inner disc. As such the tidal spiral fitting method was offset by the intrinsic patterns in the disc, so we do not attempt to fit separate regions in the wind-up diagram.

The predictions of the wind-up rates of Equation\;\ref{windEq} are show for pattern speeds of $\Omega_{\rm disc}$,  $\Omega_{\rm disc}-\kappa/2$ (the ILR) and  $\Omega_{\rm disc}-\kappa/4$ (the 4:1 ultraharmonic resonance) as dashed-dot, dotted, and dashed grey lines respectively. The gradients of each of these lines, $\dot{g}(\Omega)$, are given in the top-left of each panel to directly compare to the gradient of the fit to $\rm cot({\alpha})$.

The Flat, Rise and Dip models show spiral arms that wind-up almost exactly in accordance with a $\Omega_p=\Omega-\kappa/2$ pattern, especially at early times ($t-t_{\rm peri}<{\rm 500\,Myr}$). The Mid and Fall models follow a different trend, with wind-up rates lying somewhere between the $\Omega-\kappa/2$  and $\Omega-\kappa/4$ frequencies. The likely cause of this is the rapidly rotating inner bars of both of these models, which begin to dominate the power spectra of the $m=2$ mode for these galaxies. The FallS05 model in particular has a strong inner $m=2$ signal in the spectrograms of Figure\;\ref{spec2} very early on, which explains why the gradient in Figure\;\ref{AlphaDrop} is the furthest away from the $\Omega-\kappa/2$ model. The dominance of bars in these models also make it difficult to precisely trace the spiral structure formed by the tidal interaction, hence why there are clear breaks and scatter for the data of the Fall and Mid models in Figure\;\ref{AlphaDrop}. For example, the Mid model has a clear break to a much steeper gradient, with a transition that occurs roughly 500\,Myr after perigalacticon, equivalent to approximately 1100\,Myr since the start of the simulation. This lines up extremely well with the bar formation time (see Fig.\,\ref{PSplt} in the following section), indicating that these arms are now under the influence of the torque from the rapidly rotating inner bar. The Rise model is again the outlier, as despite exhibiting an inner bar it shows very little deviation form the ILR wind-up rate (though the minor deviation also occurs at roughly the same time as bar formation). This is likely because the bar is very slowly rotating compared to that of Fall and Mid, and so has a lesser impact on the wind-up of the tidal spirals, rotating slower than these arms at all radii.

The differences between these mass models sheds light on why the measurements of pattern speeds in simulations of tidal spirals is seemingly inconsistent between studies. The presence of inner bars, or even the underlying mode power that will form a bar much later, has the capacity to cause changes in the wind-up rate, i.e. the pattern speed of the spiral arms. This explains some studies showing tidal spirals as moving almost exactly with the ILR \citep{2015ApJ...807...73O,2016MNRAS.458.3990P} and others seeing them rotate with noticeable higher pattern speeds \citep{2000MNRAS.319..393S,2010MNRAS.403..625D} as a result of the differences (and bar-forming potential) of the different galaxy mass models.

%%%%%%%%%%%%%%%%%%%%%%%%%%%%%%%%%%%%%%%%%%%%%%%%
\subsubsection{Spur features}
%%%%%%%%%%%%%%%%%%%%%%%%%%%%%%%%%%%%%%%%%%%%%%%%
\begin{figure*}
\centering
\resizebox{1.0\hsize}{!}{\includegraphics[trim = 0mm 0mm 0mm 0mm]{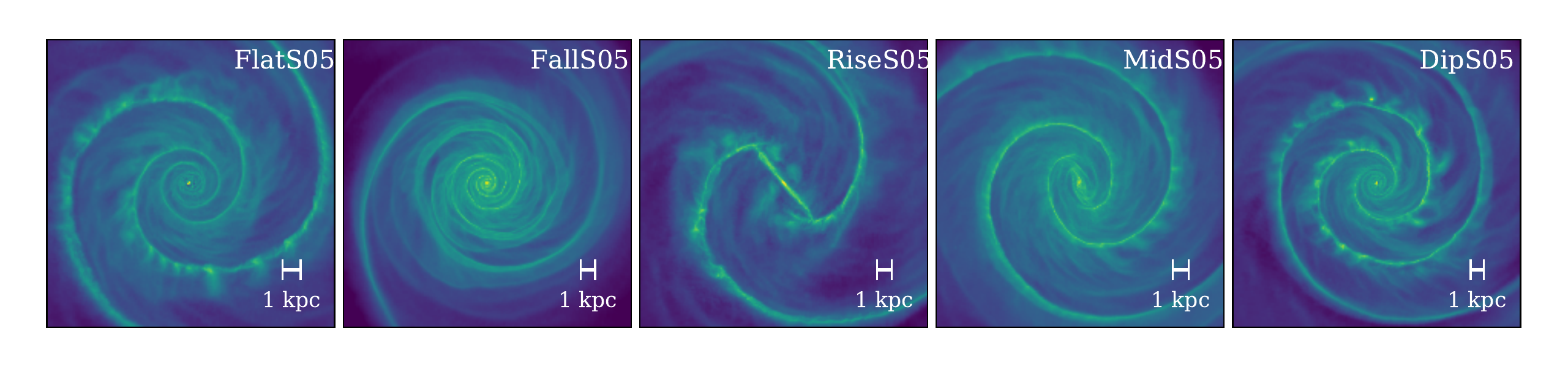}}
\caption{Spurs in the S05 family of interactions. The Flat and Dip models show the clearest spur formation, with the Fall model showing no spur generation.}
\label{GasSpur}
\end{figure*}

\begin{figure*}
\centering
\resizebox{1.0\hsize}{!}{\includegraphics[trim = 0mm 0mm 0mm 0mm]{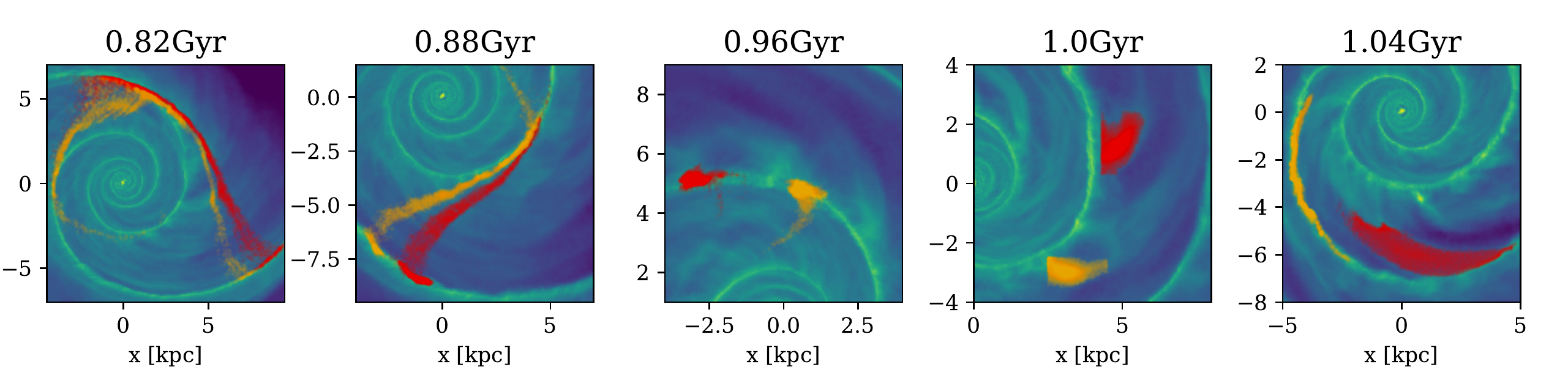}}
\caption{The location of material that forms two gaseous spurs in the DipS05 model as a function of time. The spurs are defined by the 1.0Gyr time-stamp. Earlier times clearly show their amalgamation from material at various locations in the disc, and ultimately forming two weakly bound beads located directly on the spiral arm, which are sheared out into wide azimuthal area at later times.}
\label{SpurTrack}
\end{figure*}

Spurred structures in spiral arms are observed in both observations \citep{2006ApJ...650..818L,2017ApJ...836...62S} and simulations \citep{2006ApJ...647..997S,2013MNRAS.436.1836R}. Most of the simulations investigating spurs however use fixed spiral potentials to induce spur features. Only recently has evidence of such spurs existing in tidal spiral simulations been seen \citep{2016MNRAS.458.3990P,2017MNRAS.468.4189P}, even though M51 (the poster-child of observed tidal spirals) has clear interarm spur features. In Figure\,\ref{GasSpur} we show further evidence of spurs in tidal spiral arms. The gas components of all S05 models are shown, though at slightly different epochs to highlight spur presence. All but the Fall model show distinct spur features, with the Dip model displaying the strongest. Only the bridge arm of RiseS005 displays spurred features at this time, whereas spurs in the Flat/Dip models appear in both bridge and tail arms. Interestingly, if we were to study the S10 models instead, many of these spurs become stronger, but also form clump-like structures as they leave the spiral arms.

Figure\;\ref{SpurTrack} shows the evolution and progenitor gas of two spur features in the Dip05 calculation. The five panels show different time frames, with yellow and red points showing the particles that constitute the spurs at a time of 1\,Gyr (the fourth panel across is the same timeframe used in Fig.\,\ref{GasSpur}). The two spurs are formed from an amalgamation of material from across the galactic disc, which builds up in a  arm for almost 200\,Myr before it leaves the spiral well. The spurs are then rapidly sheared away, falling into the upstream arm over a wide range of radii. It would naively appear that these spurs are simply marginally unbound clumps that form by self-gravity in spiral arms.  Credence is given to this origin by inspection of the 1\,Gyr timestamp of DipS10 in Fig\;\ref{IsoGalsE} (the stronger interaction version of the model in Fig\;\ref{SpurTrack}), which shows the clumps are sufficiently bound that they flow through the interarm regions intact. While the precise nature and origin of such spurs is the subject of debate among the community \citep{2004MNRAS.349..270W,2006MNRAS.367..873D,2006ApJ...646..213K}, this is outside the rubric of this work and we leave the in depth analysis of spurs/feathers in tidal spirals to a future study.

%%%%%%%%%%%%%%%%%%%%%%%%%%%%%%%%%%%%%%%%%%%%%%%%
%%%%%%%%%%%%%%%%%%%%%%%%%%%%%%%%%%%%%%%%%%%%%%%%
\subsection{Bar structure}
\label{sec_bar}
%%%%%%%%%%%%%%%%%%%%%%%%%%%%%%%%%%%%%%%%%%%%%%%%
%%%%%%%%%%%%%%%%%%%%%%%%%%%%%%%%%%%%%%%%%%%%%%%%

%%%%%%%%%%%%%%%%%%%%%%%%%%%%%%%%%%%%%%%%%%%%%%%%
\subsubsection{Bar morphology}
%%%%%%%%%%%%%%%%%%%%%%%%%%%%%%%%%%%%%%%%%%%%%%%%

A majority of the models here show a strong bar growth with time, and at the very least are bar unstable with underlying $m=2$ mode power. We use standard conventions \citep{2002MNRAS.330...35A,2014ApJ...790L..33L} of defining the bar via a Fourier decomposition of the stellar particles. The bar strength is often characterised by the parameter:
\begin{equation}
A_2(t) = {\rm max}\Big|_R  \left(  \frac{1}{a_0(R,t)}\sqrt{ a_2(R,t)^2 + b_2(R,t)^2}  \right)
\end{equation}
where $a$ and $b$ are the same Fourier components as in equations \ref{F_am} and \ref{F_bm} for $m=2$. We have already discussed the behaviour of this parameter as a result of bar and spiral features in Section\,\ref{sec_morph} (Fig.\,\ref{A1plt}). This parameter can also used to define bar length by finding some limiting radius where $A_2$ is below a threshold value. While this gives a good general measure of bar length, it must be used with caution as it also traces any bisymmetric structure, such as tidal spiral arms. Instead we trace the morphology of the bars by binning up particles into radial bins and then fitting Gaussian functions in polar co-ordinates. We then take the centre of the Gaussian to trace the phase of the bar feature. For any given snapshot, if the centre of the Gaussian at a given radius strays $\pm 10^\circ$ from the mean of the centroid of the Gaussian in the inner 2kpc of the disc, then the bar is deemed to have transitioned to a spiral feature. This procedure allows for the tracing of the phase and length of the bar, and is effectively the same as method v) from \citet{2002MNRAS.330...35A}. We do not attempt to fit full 2D surface density profiles to the bar (e.g. \citealt{1990MNRAS.245..130A}), which would have the added advantage of measuring a semi-minor axis.

\begin{figure}
\centering
\resizebox{1.0\hsize}{!}{\includegraphics[trim = 0mm 0mm 0mm 0mm]{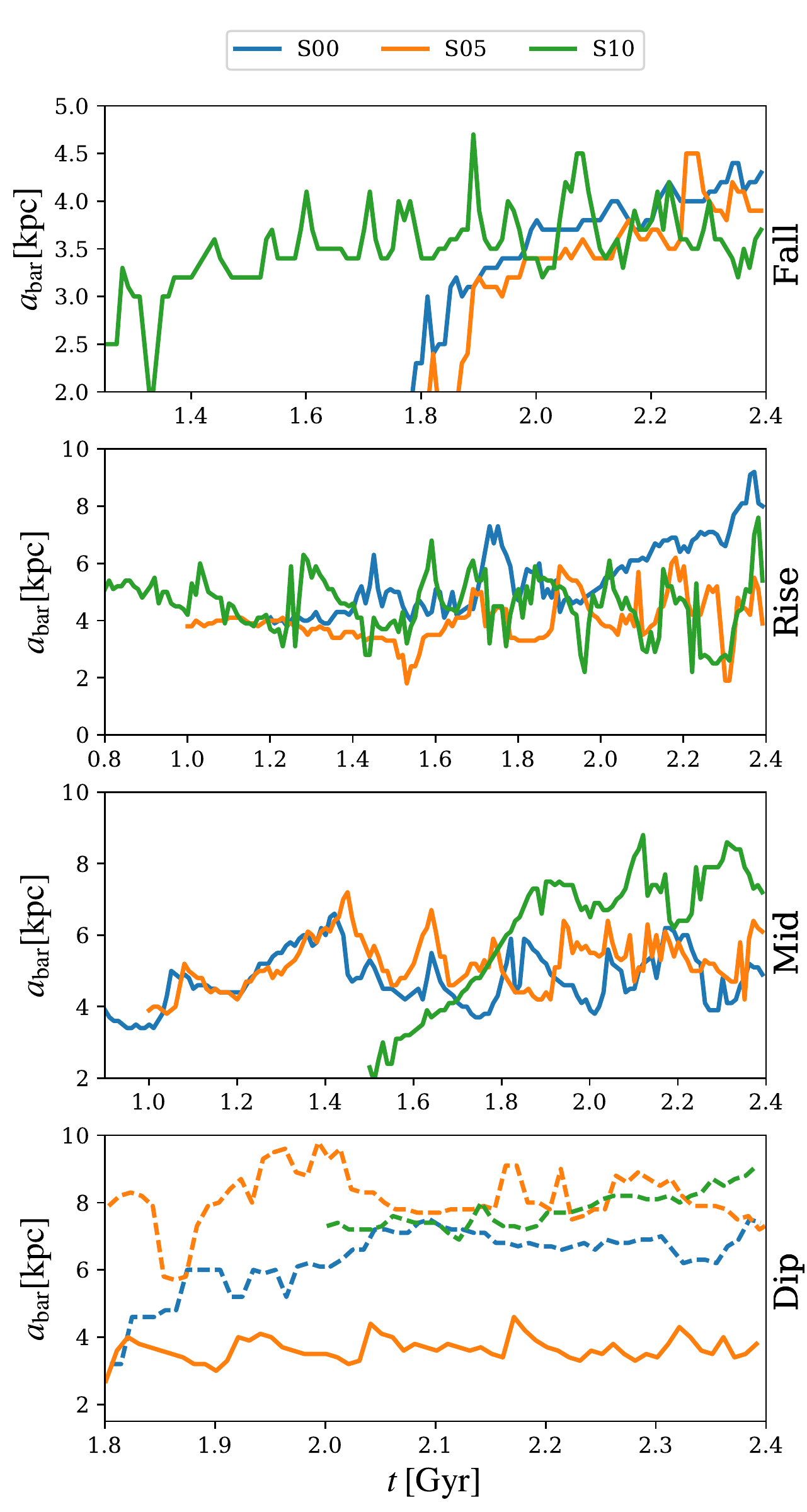}}
\caption{The bar lengths as a function of time, calculated from the direct tracing method described in the main text. The dashed lines in the Dip panel indicate bars that occur much later, and have been shifted back in time so their formation time lies within the plot (see text for details).}
\label{RBplt}
\end{figure}

We show our measurements of bar length over time in Figure\;\ref{RBplt} for all models that form clear barred features (i.e. not the Flat models). Each panel shows a different model, with different interaction scenarios shown as different coloured lines. For the Dip models, there are clearly underlying bar features that are on the cusp of appearing (see Sec.\,\ref{sec_morph} and the face-on renders in Fig.\,\ref{IsoGalsE}). We thus allowed the Dip models to evolve for an additional 5\,Gyr in order to allow bar features to manifest in the DipS00 and DipS10 calculations, and to allow the DipS05 bar to futher mature. The measurements for these late time bars are shifted into the time window of the DipS05 bar and shown as dashed lines in Figure\;\ref{RBplt}. These models have been shifted by $-6$Gyr, $-4$Gyr, and $-0.4$Gyr respectively for the S00, S05 and S10 Dip models, such that the formation time for DipS00 and DipS10 lie within the plot, while DipS05 was merely chosen at some later epoch to highlight long term evolution. Only a small window of DipS10 is shown because after the fiducial time-frame of 2.4Gyr the companion re-approaches the host galaxy, the dynamical friction being enough to bind the companion to the host, which further disrupts the disc, disguising the influence of the initial passage on the host.

In the case of the Fall model, the bars form at different times, with the strongest interaction forming a bar about 500\,Myr earlier than the other models, also consistent with the rising of $A_2$ in Figure\,\ref{A1plt}. This is the model with the strongest central concentration in the rotation curve, which has been seen to be the most difficult kind of disc to induce bar formation compared to other galaxy models \citep{1991A&A...243..118S}. The bar length seems fairly consistent across all Fall models at the end of the simulation. Simulations of bars tend to indicate that they grow in length over time (e.g. \citealt{2000ApJ...543..704D,2006ApJ...637..214M}). This bar in particular has a length of approximately $a_{\rm bar}=3.5$kpc in all cases at the end of the simulation, signifying that it is characteristic of this model galaxy, regardless of the wider environment (i.e. interactions).

The Rise model appears to have a somewhat more chaotic history in the interacting cases, where the length can change by $\pm 2$kpc over the course of the simulation. The face-on renders show that the bar ends are being stripped and distorted by the winding spiral arms, which explains why the isolated case grows in a much more steady manner than the interaction cases. Bars without central concentrations have been seen to grow (and decrease pattern speed) slower than galaxies with strong central concentrations which allow for a greater rate of exchange of angular momentum \citep{2003MNRAS.341.1179A,2015PASJ...67...63O}, implying bars formed in Rise-like models are expected to evolve slower.

The Mid model appears to behave in a different manner to the previous models. The S00 and S05 calculations are almost indistinguishable, with the face-on renders of Figure\;\ref{IsoGalsD} showing that even the bar phase is unchanged in the immediate post-interaction period. The stronger interaction, however, has the effect of suppressing bar formation, in contrast to Rise and Fall whose bars were triggered earlier by the strongest interactions. Though inspection of the face-on renders does show a small elliptical bulge-like feature in the centre of the disc of MidS10 around 1--1.2Gyr. Interactions are normally expected to induce bar formation \citep{1990A&A...230...37G,2016ApJ...826..227L}. However, if the interaction is of sufficient strength, \citet{1998ApJ...499..149M} showed it can effectively wipe-out the intrinsic bar of the disc, which appears to be the case in the Mid model. It is also possible that the large inward migration of gas in this model impedes or even destroys bar features early on \citep{1993A&A...268...65F,1998MNRAS.300...49B,2004MNRAS.347..220B}. This bar suppression does not last long however, and the bar returns much stronger than the others at later times, in line with the more accepted theory that interactions can induce stronger bar features.

The final set of models, Dip, forms a bar in the weak interaction scenario in the fiducial time-frame. This bar grows in strength over time, though it takes nearly 4Gyr to grow from 4\,kpc to 8\,kpc in length (orange solid vs dashed lines in Fig.\,\ref{RBplt}), highlighting that it has entered the secular evolution phase. The isolated case shows no strong bar features at the end of the fiducial 2.4\,Gyr simulation period, but does form a bar much later on. The strongest interaction also does not form a bar until much later, even though the gas distribution in Figure\;\ref{IsoGalsE} shows the $x_2$ orbital disc characteristic of an inner bar region. Similar to the early stages of MidS10, the stellar material in DipS10 disc may take some time to reestablish normal bar-like orbits in the wake of the wind-up of strong outer tidal spirals, compared to the S05 calculations which have experienced a quieter bar triggering by the interaction, or the bar may be being suppressed by gas infall.

%%%%%%%%%%%%%%%%%%%%%%%%%%%%%%%%%%%%%%%%%%%%%%%%
\subsubsection{Bar dynamics}
%%%%%%%%%%%%%%%%%%%%%%%%%%%%%%%%%%%%%%%%%%%%%%%%

\begin{figure}
\centering
\resizebox{1.0\hsize}{!}{\includegraphics[trim = 0mm 0mm 0mm 0mm]{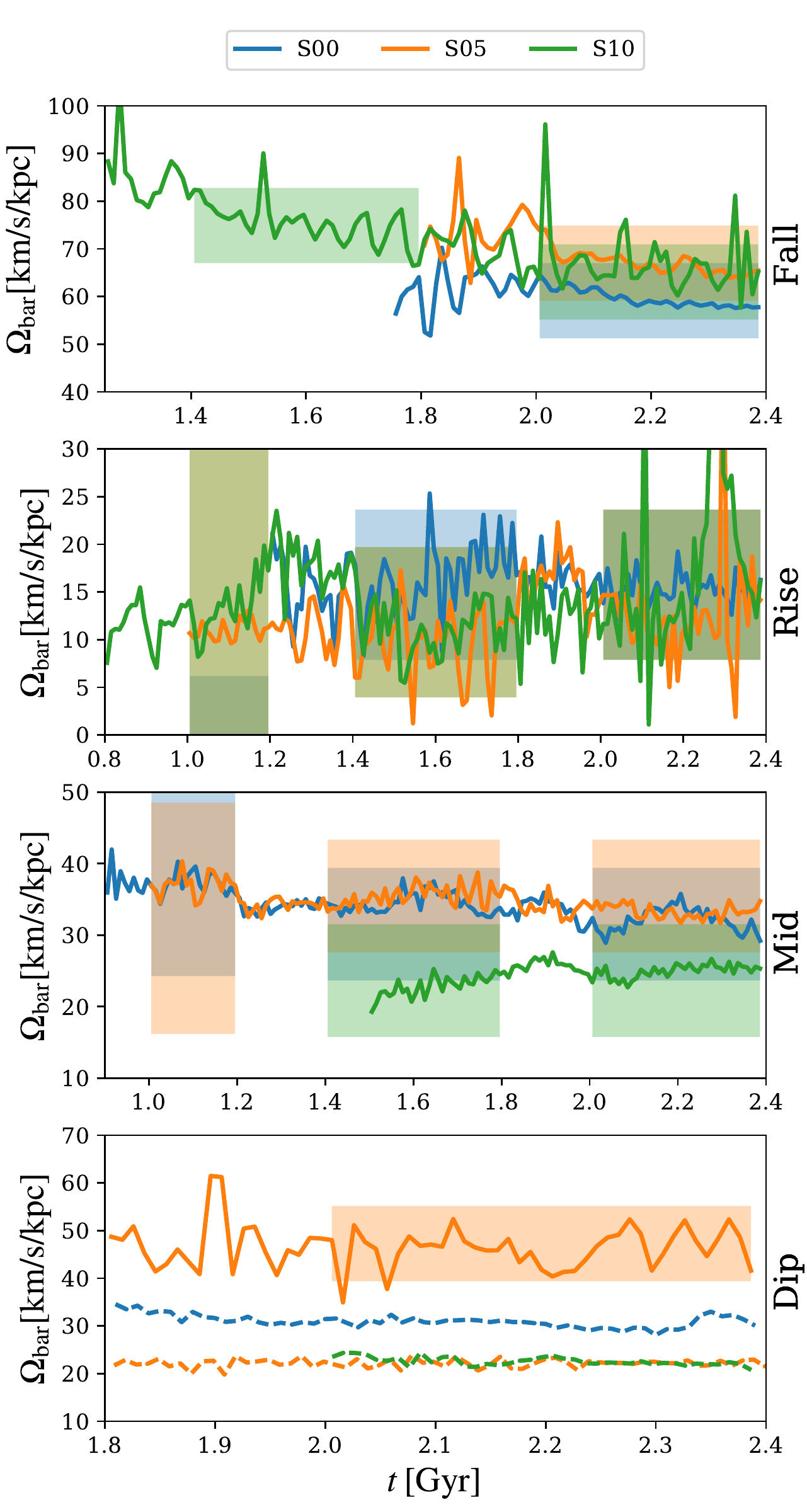}}
\caption{Bar pattern speeds as a function of time across all models where a bar is present. The solid line shows the pattern speed calculated by directly tracing the motion of stellar particles from one snapshot to the next, while the shaded boxes show the pattern speed calculated from the spectrogram analysis. The time frames for the latter are the same as those shown in Figs. \ref{spec1} to \ref{spec5}. Colours and linestyles have the same meaning as Fig.\ref{RBplt}.}
\label{PSplt}
\end{figure}

We employ two independent methods of measuring the bar pattern speed. The first is to simply measure the rate of change of the phase of the bar between snapshots, a natural by-product of measuring the bar length. The other is to isolate the horizontal ridges in our spectrogram analysis that stem from bar features. The former has the advantage of very fine temporal resolution in the changes in the bar pattern speed, though it is influenced by short term variations, such as the wind-up of outer spiral features which may throw off measurements of the bar phase. The spectrograms overcome this uncertainty, effectively averaging out small period variations, but at the cost of reduced resolution (both in time and in the precise value of $\Omega_p$ being limited by the spectrogram resolution). Both methods are thus complimentary, and we utilise both for comparison.

Figure\;\ref{PSplt} shows the values for the bar pattern speed, $\Omega_{\rm bar}$, for the bar forming simulations. The panels and colours have the same meaning as Figure\;\ref{RBplt}, with the dashed lines again indicating a late epoch for the Dip models. The lines indicate the pattern speeds as calculated using the changing phase of the bar via direct tracing, while the shaded boxes show the pattern speeds calculated via tracing peaks in the spectrogram. We chose three regions to sample from the spectrogram where possible. This includes two 0.4\,Gyr timeframes (1.4--1.8\,Gyr and 2.0--2.4\,Gyr) and a shorter timeframe of 0.2\,Gyr (1.0--1.2\,Gyr). Note that smaller time domain of the latter gives a larger uncertainty in the measured value of $\Omega_{\rm bar}$.

The first point of note is that the two different methods of pattern speed determination agree very well, with the boxed regions coinciding with the lines in all instances. The second is that there is not a huge difference between the pattern speeds between different interaction scenarios, only differing by at most $\pm 10$\ps{}. This is also seen by \citet{1990A&A...230...37G} and \citet{1991A&A...243..118S}, who find the pattern speed of bars in interactions is determined primarily by the galaxy model, rather than the properties of the interaction. Though \citet{1998ApJ...499..149M} saw that bars formed in isolation seemed to rotate faster than those triggered in interactions in some cases. The average pattern speeds seen here are around 65, 15, 30 and 45\ps{} for the Fall, Rise, Mid and Dip models respectively (in the fiducial time period). The Rise model in particular is a halo dominated galaxy, and it has been noted in previous studies that such discs are susceptible to slower rotating bars \citep{1998ApJ...499..149M}.

The Fall model appears to have slightly faster bars in the interaction scenarios, a likely by-product of the companion imparting angular momentum to these discs. The Rise model has very noisy pattern speed measurements for the interacting models, with values broadly equivalent regardless of the presence of a perturber. This is somewhat surprising given that the companion clearly influences the bar, by triggering it much earlier in S10 than in the isolated case. For the Mid model the pattern speed only shows a change for the strongest interaction (as with $a_{\rm bar}$), with the S00 and S05 models showing a gradual decrease in pattern speed over time. The bar in the early epoch of Dip05 does not change in the 0.6\,Gyr window shown, but does gradually decrease over the course of 5\,Gyr, reaching as low as 20\ps{} in the later epoch (dashed orange line). 

There is only minimal changes in pattern speeds of bars over time, though the late epoch of the DipS05 model and the FallS10 model hint that slow down is indeed taking place, but just on a much longer timescale than that shown here. This may indicate these bars have not yet have truly entered the secular evolution slow-down phase (e.g. \citealt{2013seg..book....1K}). Tidally induced bars in published simulations also tend to have slowly decreasing pattern speeds over the period of a few Gyr \citep{1991A&A...243..118S,1998ApJ...499..149M,2017MNRAS.464.1502M}.

\begin{figure}
\centering
\resizebox{1.0\hsize}{!}{\includegraphics[trim = 0mm 0mm 0mm 5mm]{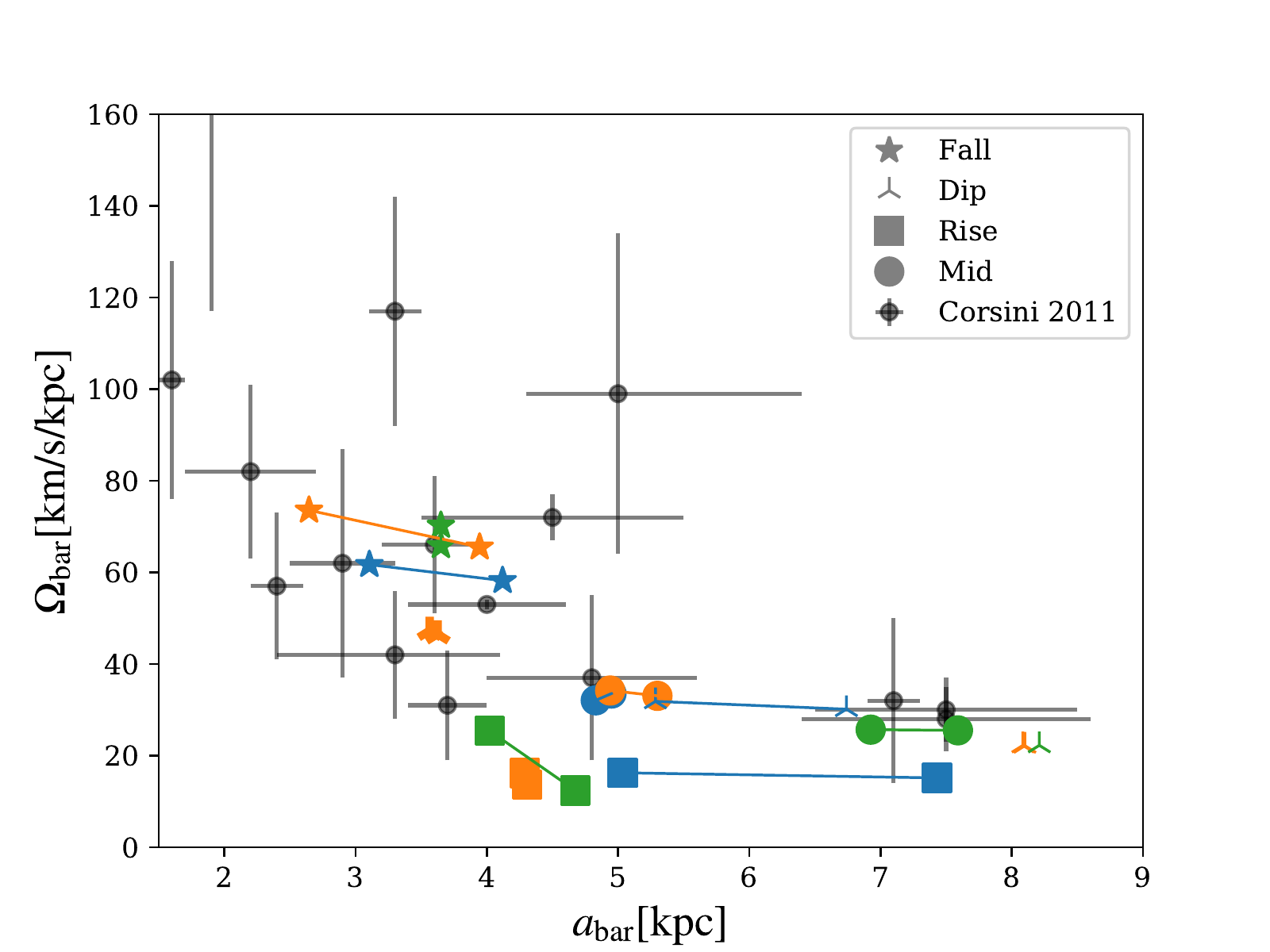}}
\caption{Mean values for bar length plotted against pattern speed for the time frame of 1.8 to 2.0 Gyr, and 2.2 to 2.4Gyr, connected by solid lines. The non-bold symbols for Dip show the bar patten speed at a much later epoch. Observational data from \citet{2011MSAIS..18...23C} is shown as the black points.}
\label{RBpltObs}
\end{figure}

\begin{figure}
\centering
\resizebox{0.9\hsize}{!}{\includegraphics[trim = 0mm 0mm 0mm 0mm]{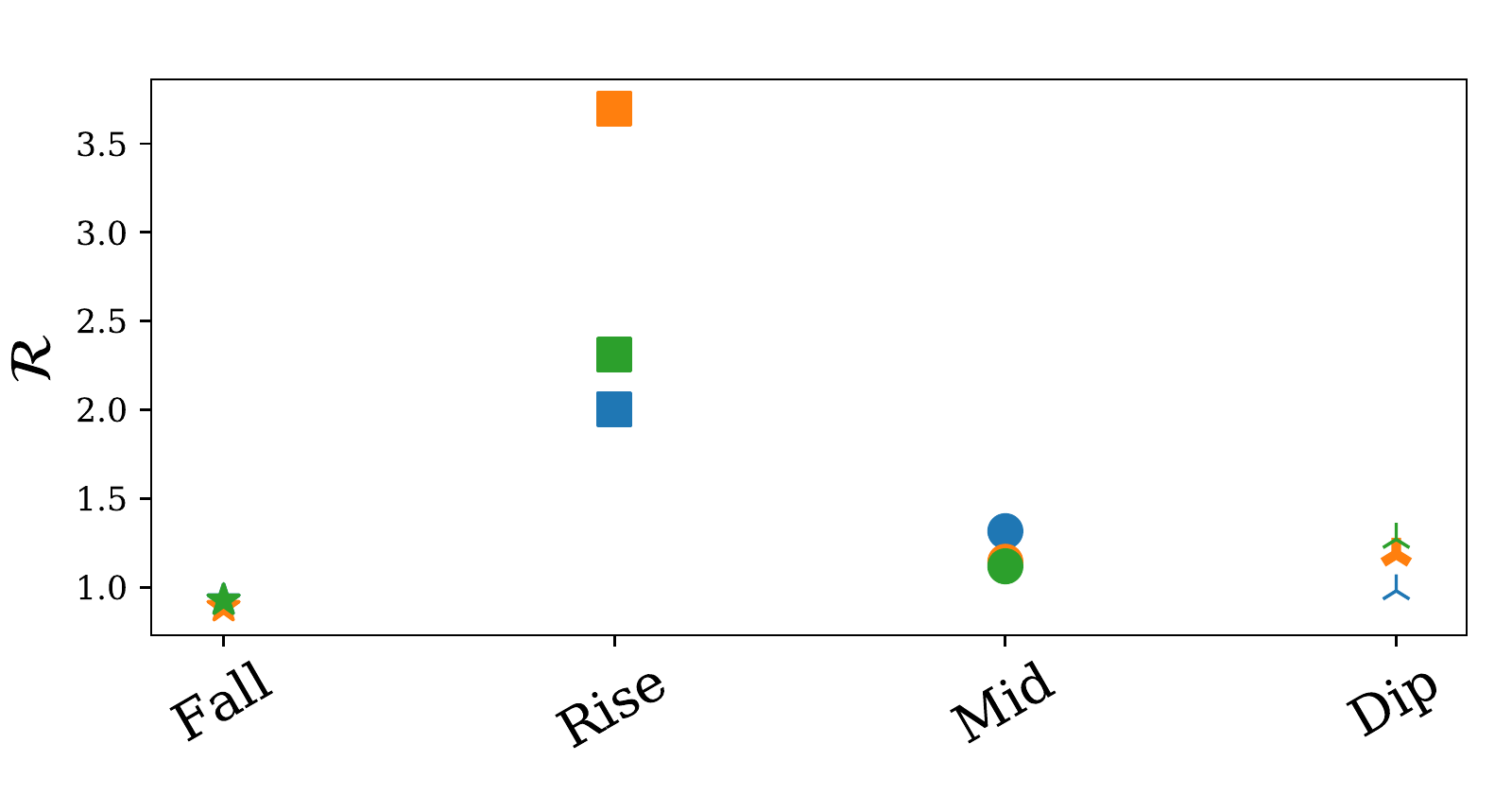}}
\caption{The $\mathcal{R}$ parameter for all bars formed in this study. The Fall, Mid and Dip bars are fast rotators, while Rise displays slow bars.}
\label{Rparam}
\end{figure}

The $\mathcal{R}$ parameter is used to describe the bar rotation in relation to its length via:
\begin{equation}
\mathcal{R}\equiv \frac{R_{\rm CR}}{a_{\rm bar}}
\end{equation}
where $R_{\rm CR}$ is the co-rotation radius. Theory predicts this parameter has values in the range of $\mathcal{R}=1.2\pm 0.2$ \citep{1992MNRAS.259..345A,2000ApJ...543..704D}. Bars are generally designated as ``fast" ($1.\leq \mathcal{R}\leq 1.4$) or ``slow" ($\mathcal{R}> 1.4$) rotators, with observed bars tending to be ``fast" rotators
\citep{2008IAUS..245..125C,2009ApJ...704.1657F,2011MSAIS..18...23C,2015A&A...576A.102A}. High values of $\mathcal R$ (2--3) are seen in some simulations of interacting galaxies \citep{2016ApJ...826..227L,2017ApJ...842...56G}, a small number of isolated simulations \citep{2008MNRAS.388.1803R} and individual observed galaxies \citep{2009A&A...499L..25C}. Having $\mathcal{R}< 1.0$ is assumed to be impossible, resulting in bar dissipation. However, such low values have been seen in both observations and simulations \citep{2003MNRAS.342.1194D,2015A&A...576A.102A,2017MNRAS.469.1054A}.

In Figure\;\ref{RBpltObs} we plot the mean pattern speed for our bars against the bar length, compared to the observed data from \citet{2011MSAIS..18...23C}, and in Figure\;\ref{Rparam} we show the $\mathcal{R}$ parameter for the simulation data. For Figure\;\ref{RBpltObs} we show two different times; 1.8--2.0Gyr (where possible) and 2.2--2.4Gyr, connected by solid lines (this later epoch is used to calculate the $\mathcal{R}$ values in Fig.\;\ref{Rparam}). Generally the bars formed here trace the same region of parameter space as the observed galaxies, with the smallest bars rotating the fastest. The galaxies with the largest disc to halo mass ratio (Fall) have the highest rotation speed, consistent with previous works \citep{1987MNRAS.228..635N}. The solid lines clearly show bars slowing down and growing over time. The dearth of very short bars with rapid rotations is because our rotation curves have been normalised to 200$\rm km\,s^{-1}$ for consistency, and so are biased against very high orbital velocities. For example, NGC\,2950 is a point in the upper left of this plot with a bar rotation speed of around 120\ps{}, but a circular velocity on the order of 360${\rm km\,s^{-1}}$ \citep{2003ApJ...599L..29C}.

\begin{figure*}
\centering
\resizebox{1.0\hsize}{!}{\includegraphics[trim = 0mm 0mm 0mm 0mm]{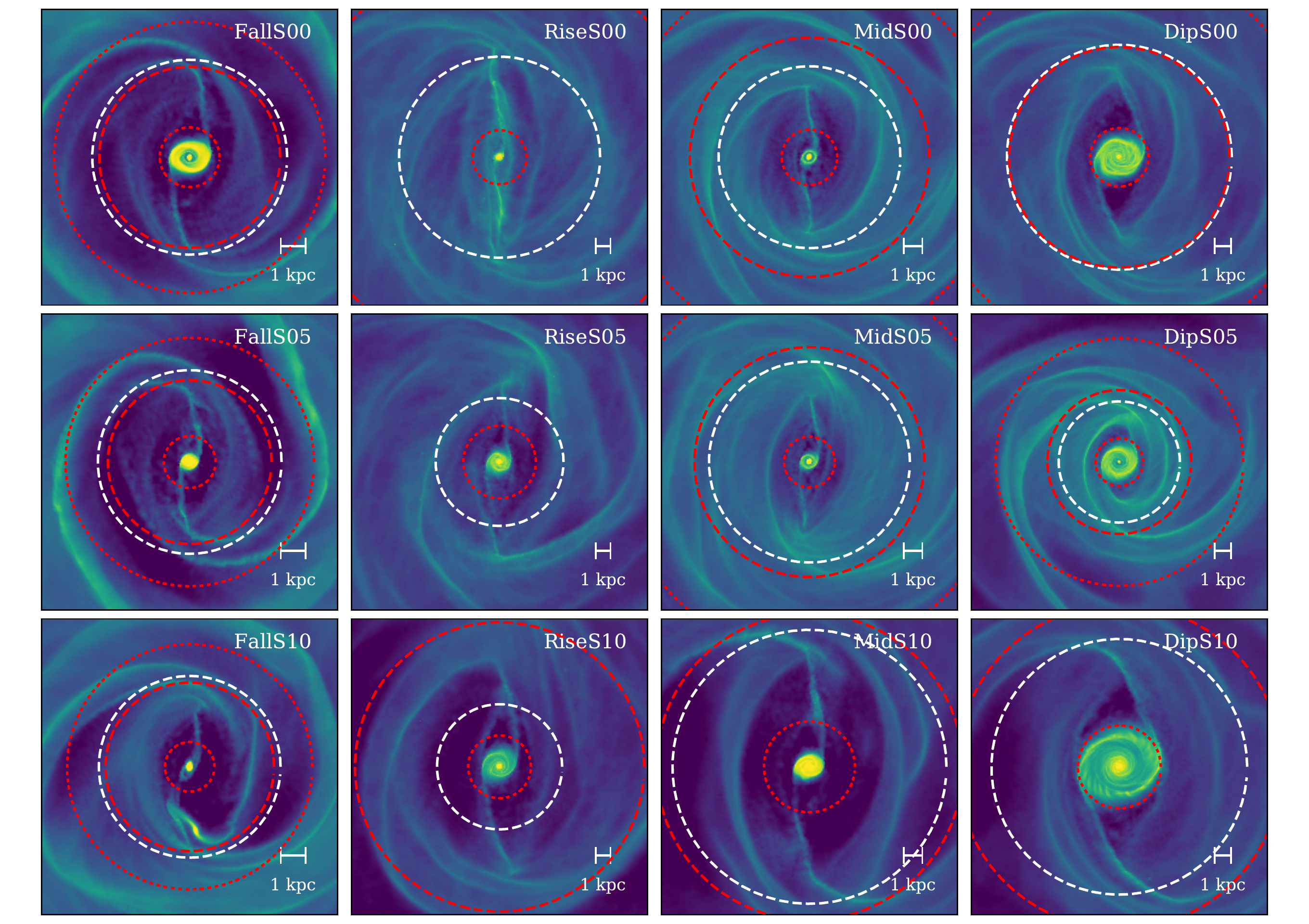}}
\caption{The gas structure in all bars formed in the simulations presented here. All are shown at a time of 2.4Gyr, except for Dip, which is shown at a later epoch. The white dashed line shows the bar length as determined by direct particle tracing. The red dashed line shows the radius of co-rotation, using the pattern speeds also determined by the particle tracing. Dotted red lines show the inner and outer Lindblad resonances, the latter of which is only visible in a few of these panels.}
\label{GasBar}
\end{figure*}

There is no clear correlation in phase space as to where the different interaction scenarios lie, i.e. if S10 drives an increase in bar length for example. Bars formed within the Fall, Mid and Dip calculations appear to be ``fast" rotators, with $\mathcal{R}< 1.4$. The Rise simulations however, appear as ``slow" rotators, with RiseS05 in particular having $\mathcal{R}>3$. The interaction in RiseS05 has caused a disconnect between an outer spiral and inner bar region (see the face-on maps of Fig.\,\ref{IsoGalsC}). As the spirals wrap around the disc they effectively eat away at the bar ends, impeding bar length to below the level of RiseS00. This also occurs in RiseS10, but the bar has been triggered much earlier and stronger in this model, allowing it outgrow the RiseS05 bar. This goes some way to explaining the high values of $\mathcal{R}$ in Rise. Bars with $\mathcal{R}>2$ are also seen in the interaction simulations of \citet{2017ApJ...842...56G}. However, even with this underestimated bar length, the Rise models are still characterised by slow bars, evident from RiseS00 still displaying $\mathcal{R}\approx 2$.

To further highlight the significance of high values of $\mathcal{R}$, we plot the gas density around the bar regions for each barred simulation in Figure\,\ref{GasBar}. Different models are shown in different columns and different rows show different interaction scenarios. All bars are orientated vertically, with the bar ILR, CR and OLR regions indicated by the red dotted, dashed and dotted circles (in that order from the centre). The white dashed circle shows the bar length as calculated for Figure\,\ref{RBplt}. Each column shows a different spatial scale, but the same scale is used across the same mass models.

It is clear from this plot which are fast and which are slow rotators. Fall, Mid and Dip all generate fast bars with $\mathcal{R}< 1.4$, as seen by how close the co-rotation radius is to the bar lengths (white and red dashed lines). Conversely, Rise models are slow bars, with the co-rotation radius often lying off panel. These face-on maps also highlight the differences between bar structures, with all having very clear dust-like lanes that are fuelling the inner bar region. These connect from the bar ends to the inner $x_2$ orbital disc, with some appearing almost perfectly straight (Rise00). 

The $x_2$ gas discs are not the same across similar models. In the case of Rise, Mid, and Dip, the size of the disc is largest for the strong interactions (see also the 2.4\,Gyr images of Flat in Fig.\,\ref{IsoGalsA}). These discs are always contained within the ILR. The Dip models in particular show an excellent correlation between the $x_2$ region and the ILR, which is likely the result of the Dip models having the strongest inner bulge component. The growth of these regions can be attributed to material infall induced by in the interactions. As for the Fall model, we postulate that the nature of the interaction has given the outer/mid disc enough angular momentum to halt the infall of material into the disc centre at the same rate as FallS00. This galaxy, with its falling rotation curve, has the greatest shear in the disc \citep{nganpaper}, which may be efficient in acting with the imparted angular momentum into the outer disc to stop the infall of material that otherwise would have occurred in the other galaxy models. The detailed analysis of the migration/infall of material in tidal bars is one we leave to a future work.

%%%%%%%%%%%%%%%%%%%%%%%%%%%%%%%%%%%%%%%%%%%%%%%%
\subsubsection{Angular momentum transfer}
%%%%%%%%%%%%%%%%%%%%%%%%%%%%%%%%%%%%%%%%%%%%%%%%

\begin{figure*}
\centering
\resizebox{0.9\hsize}{!}{\includegraphics[trim = 0mm 0mm 0mm 0mm]{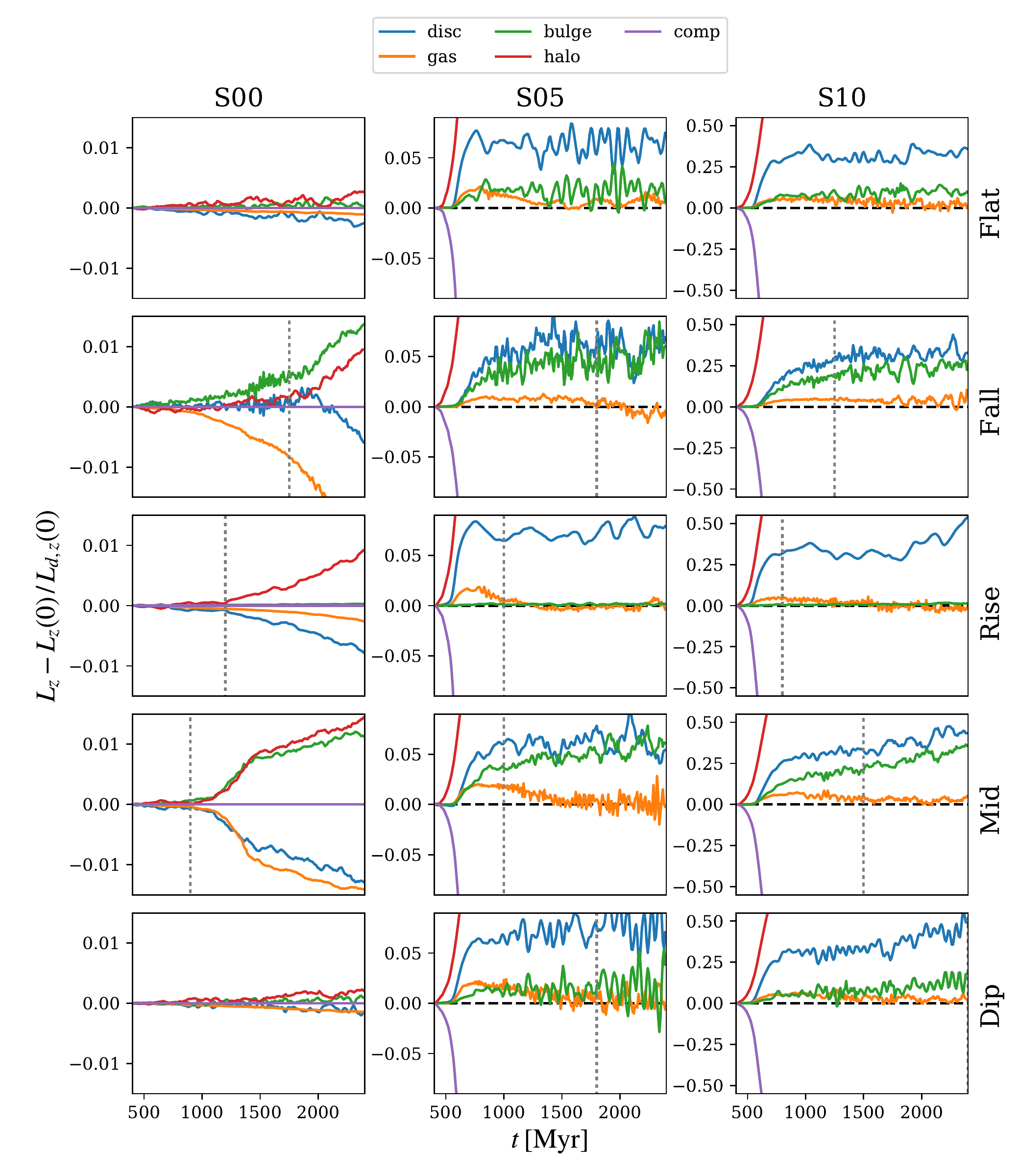}}
\caption{Changes in the angular momentum $L_z$ for all particles over time in the 15 simulations presented here. The vertical axes denotes the difference compared to $L_z$ at $t=0$ as a fraction of the initial disc angular momentum: $L_{d,z}(0)$. Different mass models are shown in each row and different interaction scenarios in each column. Each coloured line indicates a different particle type, with the dashed horizontal line showing the $\Delta L_z=0$ line. The vertical dotted line indicates the bar formation time where relevant.}
\label{Lzplot}
\end{figure*}

The change in angular momentum in each simulation is calculated in an attempt to identify the differences in bar structure in each simulation. Figure\;\ref{Lzplot} shows the change in the value of $L_z-L_z(0)/L_{d,z}(0)$ as a function of time for each of the 15 simulations presented in this work, where $L_z$ is the $z$-component of angular momentum, $L_z(0)$ is the value at the start of the simulation, and $L_{d,z}(0)$ is the initial angular momentum of the disc (used as a normalisation term). In each panel we show the amount of angular momentum contained within each component: disc, gas, bulge, halo and companion. The vertical dotted lines show the approximate bar formation time (the same as used in Fig.\;\ref{A1plt}). 

For the isolated discs the three bar forming galaxies (Fall, Rise, Mid) all have steep rises in angular momentum transfer over time, with the discs transferring angular momentum to the axisymmetric bulge and halo components. The bar-free discs (Flat and Dip) have a much lower transfer rate, though still have a noticeable net gain of halo angular momentum. It has been well document in past numerical studies that bars are characterised by a transfer of angular momentum from the disc to the halo component \citep{2002ApJ...569L..83A,2003MNRAS.341.1179A}, which is supported by our results here. However, the angular momentum transfer seems to be ahead of the primary bar formation in the Fall model. This is due to a very small, rapidly rotating inner nuclear bar. This is too small to be detected by our bar-finding routine, and appears kinematically distinct from the much larger bar formed at later times. The Rise model has a steady angular momentum exchange, indicating a more gradual bar formation rate. The Mid model has a minor angular momentum transfer phase before bar formation, and then a rapid increase in the transfer from both discs to the halo and bulge components after the bar has been formed, only to then slow down again about 500\,Myr later.

All of the interaction simulations paint a very similar picture regardless of the host galaxy mass model. The majority of the angular momentum transfer is between the companion and the halo, and is not shown to full scale in the figure to instead focus on the change in the discs. All discs seem to gain some angular momentum from the companion. The gas discs only see a slight increase, and eventually impart some of their momentum back to other components. In the case of S05 the gas angular momentum has suffered a net loss by the end of the simulation for all but the Flat simulation (the only clearly bar-free model), suggesting this is due to transfer of angular momentum to the bars as gas migrates inwards. This is less clear in the stronger interaction models (S10) and only the Rise model shows the gas suffering a net angular momentum loss.

The magnitude of angular momentum exchange is clearly the strongest in the S10 interactions, followed by the S05 and then S00 calculations. The bars in the S00 models only transfer about 0.01 of the angular momentum from the discs to the axisymmetric components, whereas when a companion is introduced amount of angular momentum being transferred increases by orders of magnitude. As such, it is difficult to disentangle the angular momentum transfer that is fuelling the bar formation from what is simply inherent to the interaction \citep{2017ApJ...842...56G}.

%%%%%%%%%%%%%%%%%%%%%%%%%%%%%%%%%%%%%%%%%%%%%%%%
%%%%%%%%%%%%%%%%%%%%%%%%%%%%%%%%%%%%%%%%%%%%%%%%
\subsection{Comparison to observed morphologies}
%%%%%%%%%%%%%%%%%%%%%%%%%%%%%%%%%%%%%%%%%%%%%%%%
%%%%%%%%%%%%%%%%%%%%%%%%%%%%%%%%%%%%%%%%%%%%%%%%

\begin{figure*}
\centering
\resizebox{0.9\hsize}{!}{\includegraphics[trim = 0mm 0mm 0mm 0mm]{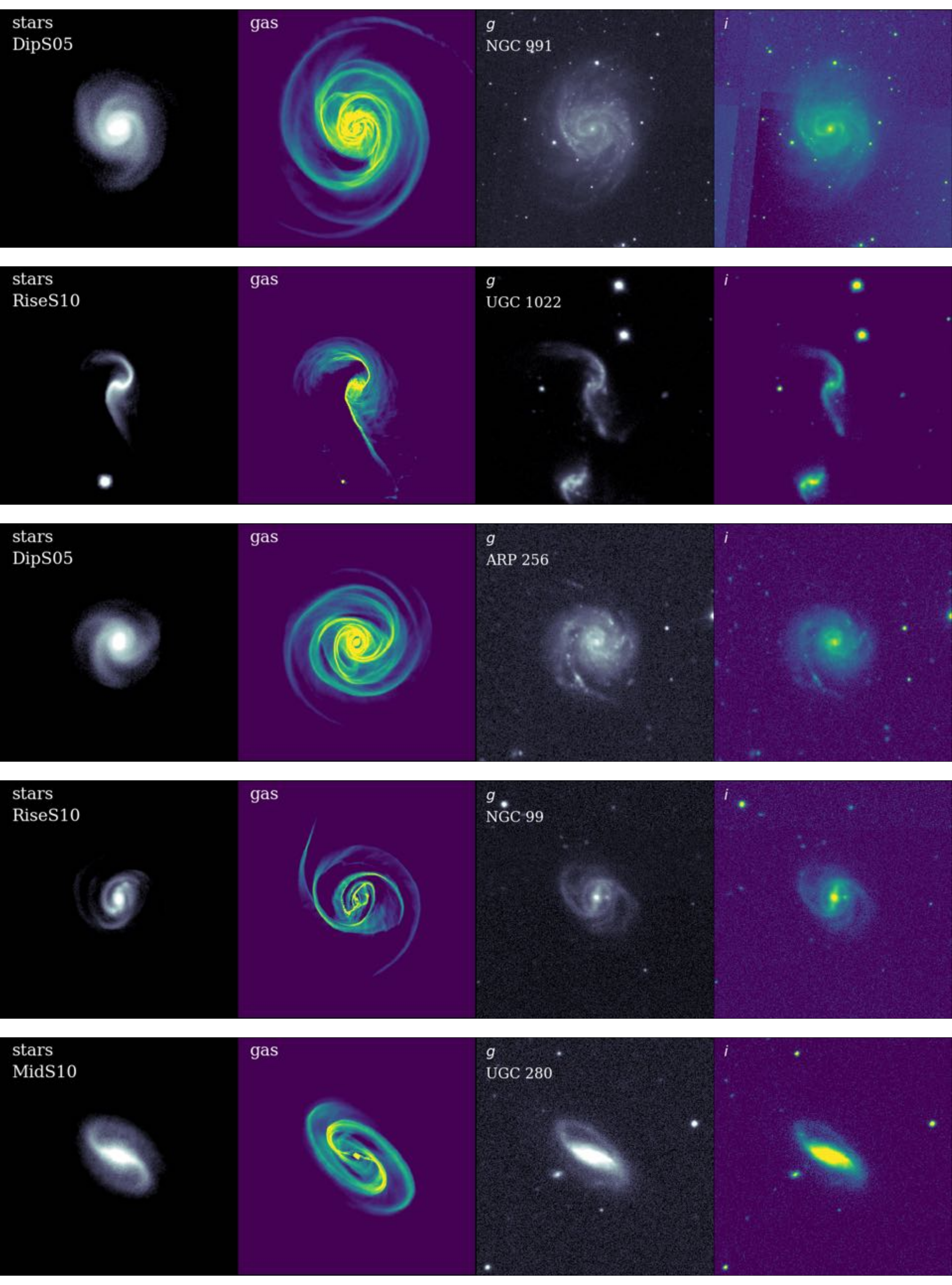}}
\caption{A selection of simulated galaxies compared to observed discs. The first and second columns show the simulated stellar distribution and gas column density. Gas surface densities are shown on in the range of $0.5\,{\rm M_\odot\,pc^{-2}}\leq \Sigma_{\rm g}\leq 10{\rm M_\odot\,pc^{-2}}$ and stellar surface densities in the range $8\,{\rm M_\odot\,pc^{-2}}\leq \Sigma_{\rm *}\leq 500{\rm M_\odot\,pc^{-2}}$. The third and fourth columns show the $g$ and $i$ band SDSS magnitudes of galaxies from the EFIGI catalogue \citep{2011A&A...532A..74B} with similar morphologies.}
\label{obs1}
\end{figure*}

\begin{figure*}
\centering
\resizebox{0.9\hsize}{!}{\includegraphics[trim = 0mm 0mm 0mm 0mm]{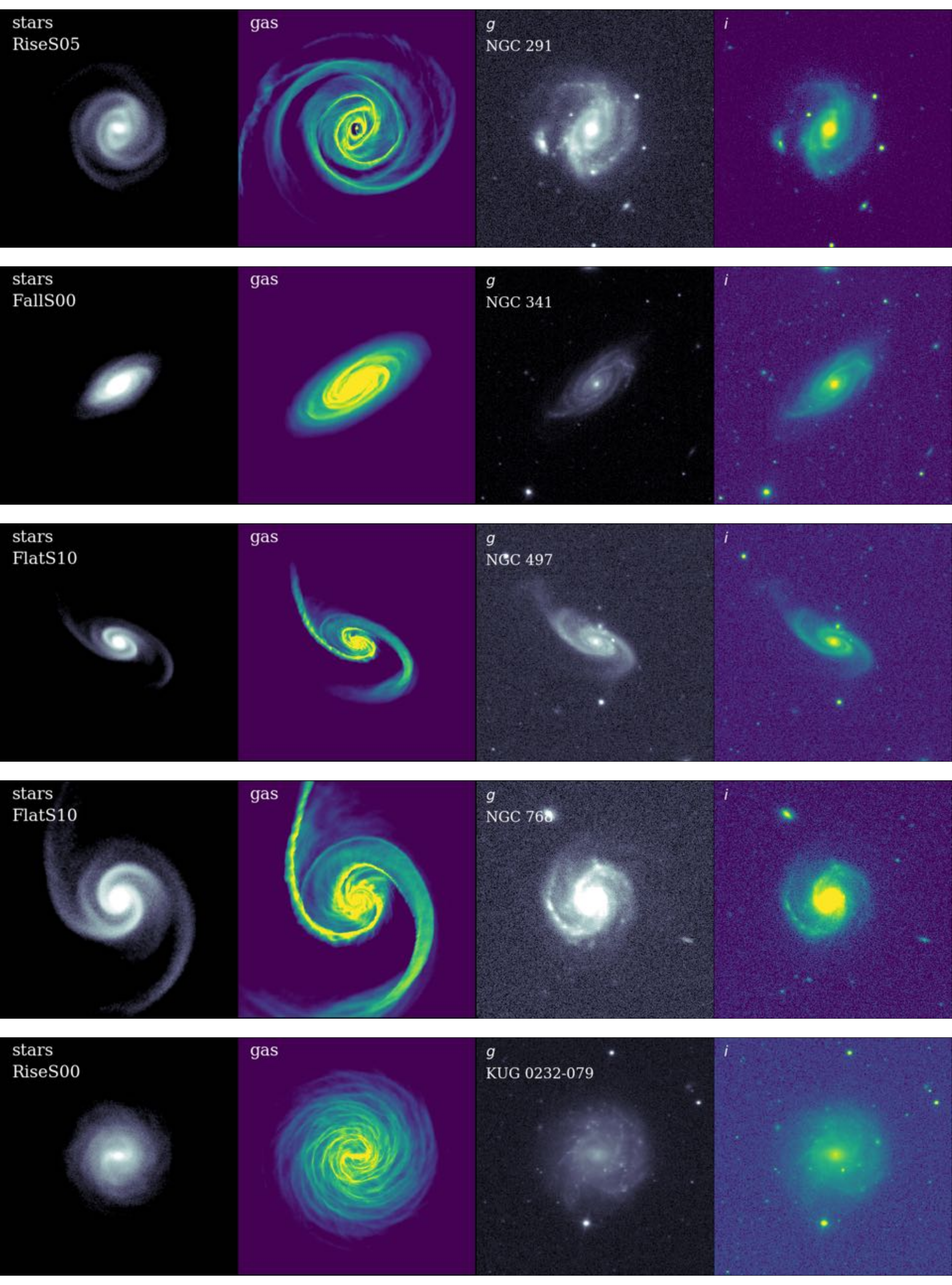}}
\caption{As Fig. \ref{obs1} for 5 additional observed galaxies.}
\label{obs2}
\end{figure*}

The galaxies produced in this study create a wide variety of different galactic structures. Such a catalogue would be a useful resource to determine the origin of the morphology of observed galaxies. Figures \ref{obs1} and \ref{obs2} are a proof of concept, showing a selection of 10 observed galaxies alongside similar simulation outputs from this work. These figures are meant to only provide a qualitative comparison, having simply been chosen by-eye by matching the stellar component to the $g$-band images, and we leave a quantitive fitting procedure to a future study. The observed galaxies are taken from the EFIGI database \citep{2011A&A...532A..74B}, which catalogues 4458 galaxies form the Sloan Digital Sky Survey (SDSS) in 5 different bands. We show galaxy data in the $g$ and $i$ bands, using them as proxies for the stellar and gaseous distributions in our simulations. These galaxies were simply selected from the first 260 in the catalogue as having clear spiral or bar structure, with low inclination on the sky, and features that matched our simulations. There are a great many other galaxies in the catalogue that provide good matches to other simulation snapshots, but performing such a comparison would require some automated fitting procedure, and this is outside of the goal of this paper to perform such analysis. The galaxies are identified by their NED names which are (from top to bottom in the Figure \ref{obs1}): UGC\,10122, ARP\,256, NGC\,99, UGC\,280, NGC\,291 and (in Figure\,\ref{obs2}): NGC\,341, NGC\,497, NGC\,768, KUG\,0232-079, NGC\,991.

We show gas and stellar surface densities for our simulation snapshots in the range of $0.5\,{\rm M_\odot\,pc^{-2}}\leq \Sigma_{\rm g}\leq 10{\rm M_\odot\,pc^{-2}}$ and $8\,{\rm M_\odot\,pc^{-2}}\leq \Sigma_{*}\leq 500{\rm M_\odot\,pc^{-2}}$. The gas scaling was chosen to fall in the range of HI gas in observed galaxy surveys such as THINGS \citep{2008AJ....136.2846B}. Inspection of our gas maps show that surface density is often above this threshold. However, to properly incorporate this density regime requires some treatment of molecular gas column densities and thus ISM chemistry, as well consideration of the star formation process eating away at the densest regions. As such, we chose this somewhat crude density threshold for our gas maps.

Our synthetic galaxy sample reproduces several structures seen in these external galaxies, including both spiral and barred features. Out of these matches, two are well reproduced by galaxies in isolation, while the other 8 are better matches to interaction scenarios. Three of these (UGC\,1022, NGC\,497 and NGC\,768) are only shortly after perigalacticon passage with the companion, matching the strong $m=2$ outer features seen in the observed discs. UGC\,1022 in particular even has the companion in the simulation placed at the same position as the satellite galaxy in the observed data (at the bottom of the panels). The gas is not as good a match to $i$-band images as the stellar data is to the $g$-band, as the $i$-band is a poorer tracer of gas than traditional radio sources. The gas discs in our models extend well beyond the stellar discs, similar to what is seen in external HI surveys \citep{2008AJ....136.2563W,2009AJ....138.1741C}, though this is expected by design of the initial conditions. Finding non-axisymmetric gas features at such extended distances could provide evidence of some tidal passage, such as those seen in M51 \citep{1990AJ....100..387R} and M81 \citep{1994Natur.372..530Y}. For example, NGC\,99 in Figure\,\ref{obs1} is matched to a simulation that shows clear outer spirals produced in the companion passage, highlighting the importance of future radio surveys in identifying the origin of spiral arms.

In the 260 galaxies from the EFIGI catalogue we inspected, there is a family of barred-spirals that we found impossible to match with our simulations. These SB(r)b/SB(r)bc (e.g. UGC\,280, NGC\,151, UGC\,719) type galaxies have clear inner rings surrounding their bars but with spiral features seemingly disconnected. It may be possible that such galaxies have suffered an interaction after the formation of their bars. For example, the Fall models form short bars after about 2\,Gyr, creating ring-like structures in the gas, but this is long after the companion has passed (though FallS10 has a hint of an inner bar/ring and disconnected outer spirals). Inducing companions into such models after many Gyr of isolated evolution would likely create such features lacking in our sample.

A more rigorous fitting procedure would shed light on the origin of specific spiral features \citep{2010MNRAS.403..625D}, aiding in determining bar patterns speeds \citep{2008MNRAS.388.1803R}, and help identify possible dark nearby companions that have induced bar/spiral features in observed galaxies \citep{2011ApJ...743...35C}. Note that the addition of rotation curve data would aid in finding more robust matches than simply matching morphology alone, though this would significantly restrict the observational sample size.

%%%%%%%%%%%%%%%%%%%%%%%%%%%%%%%%%%%%%%%%%%%%%%%%
%%%%%%%%%%%%%%%%%%%%%%%%%%%%%%%%%%%%%%%%%%%%%%%%
%%%%%%%%%%%%%%%%%%%%%%%%%%%%%%%%%%%%%%%%%%%%%%%%
\section{Conclusions}
\label{sec:conc}
%%%%%%%%%%%%%%%%%%%%%%%%%%%%%%%%%%%%%%%%%%%%%%%%
%%%%%%%%%%%%%%%%%%%%%%%%%%%%%%%%%%%%%%%%%%%%%%%%
%%%%%%%%%%%%%%%%%%%%%%%%%%%%%%%%%%%%%%%%%%%%%%%%
In this work we have performed a number of simulations of disc galaxies in different interaction scenarios. Five distinct mass models were considered, based on rotation curve data from observed galaxies. For each mass model we considered three interaction scenarios; isolation, weakly perturbed, and strongly perturbed by some passing satellite. The purpose was to assess and quantify changes in the spiral and bar features that may result from the induced tidal forces.

The interactions induced clear spiral features in each of the stellar and gaseous discs, with some altering only the outer disc while others extend deep into the galactic centre. The stronger interactions tended to create more chaotic and disturbed discs, while weaker interactions created smooth logarithmic spiral patterns. The more centrally concentrated rotation curves tend to have a limited response in the inner disc, with the greater inner mass concentration impeding the propagation of spiral waves into the centre. Spirals appear strongest in rotation curves with a minimal central concentration, with orbits changing direction by almost 90$^\circ$ as they encounter the spiral front. Such rotation curves are seen in dark matter dominated systems and dwarf galaxies with increasing rotation with radius.

By tracing the winding rate of the tidal spirals over time we find that in three out of five of our models they rotate with a pattern speed very close to the 2:1 orbital frequencies. For the remaining models, with higher central concentrations, such features can be triggered into moving faster by the appearance of bar features, causing them to appear to wind much faster (though still slower than the material speed). This is a possible explanation of why there is some discrepancy between literature studies as to the rotation speed of tidal spirals, with underlying bar modes caused by intrinsic differences in galaxy mass models influencing the evolution of tidal arms. A spectrogram analysis reveals a similar rotational behaviour, and clearly shows the transition of the tidal spirals into features of constant pattern speeds (i.e. bars). Interactions can also imprint longer-lived 2-armed spiral features to appear in discs well after the passing of the companion where they would otherwise not show such features, with all discs showing some strong constant pattern speed feature Gyr's after perigalacticon passage. The isolated Milky Way-like model (dipped rotation curve) in particular shows pattern speeds that are an excellent agreement with those of the inner (bar dominated) and outer (spiral dominated) regions of the Galaxy, but only in the isolated and weak interaction cases.

Three out of five of our models show spirals with clear spur-like features on the convex side of the spiral arm. These are seen in discs with the flattest of rotation curves, with halo or bulge/disc dominated systems forming very little or no spur features. This is one of only a few instances in the literature where spurs are seen in tidal spirals, without the need for a density wave potential of constant pattern speed.

All of the discs presented are bar unstable, and the interactions seem to have some noticeable impact on the properties of the bars. Bars formed in isolation appear to stem from features rotating with the 4:1 frequency of the disc, whereas in interactions they appear to grow from features rotating with the 2:1 resonance (i.e. the tidal arms). Three of the five models show clear bar features formed in the time frame of the simulations, and in two of those the stronger interactions induce bars much faster than the isolated and weak interactions (the dark matter and disc+bulge dominated rotation curves). In the other case (where the rotation curve peaks mid-disc) the bar formation is delayed by approximately 500\,Myr in the strongest interaction compared to the isolated case, though once the bar forms it is stronger than in the other two cases. This appears due to the collapse of the gas disc into clumps, triggered by the companion passage. These disrupt angular momentum transfer to stellar bar until they migrate to the galactic centre, then resulting in the creation a larger bar than in the weak interaction and isolated case. For the two other models with effectively flat rotation curves, the companion appears to have induced early-stage bar features where none yet existed in the isolated cases, with the stronger interactions seeding larger bars, evident from the extent of the $x_2$ orbital features and arms extending form the bar end seen in the gas. 

In summary: interactions slightly accelerated bar formation in two models (dark matter and disc+bulge dominated curves) with the bar relatively unchanged, induced bar features in flat rotation curve discs where none existed before, and delayed bar formation in the disc-dominated rotation curve model likely due to interference from gas clumping. The preference for bars to be formed earlier in interactions than in the isolated cases agrees with that seen by other studies (e.g. \citealt{1990A&A...230...37G}).

Of all the bars formed, most appear to be in the ``fast" category ($1.0<\mathcal{R}<1.4$), with only the dark matter dominated curves producing ``slow" rotators. There is no noticeable correlation between the bar lengths or pattern speeds and the strength of an interaction. We concur with the interpretation of \citet{2017A&A...604A..75M} that there may be no grand universal picture of how interactions impact bar dynamics and shape, with the results being a complex result of the mass ratios, gas content, perturber properties, and even the phase angle of asymmetric features in the primary disc.

The models shown here display a wide variety of different morphological features, ranging from short bars, wide angled grand design spirals, tightly wound arms, flocculent discs, arm-bar disconnects and irregular outer arm features. We show that such a catalogue can easily reproduce a range of features seen in real galaxies. This suggests that tidal interactions with small passing satellites such as dark matter subhaloes could be responsible for much more of the observed galactic morphologies than naively expected, even when the companion has long left the system. In a future work we aim to embark on an automated reproduction of a large population of observed galactic discs, shedding light on the key drivers of their morphological features.

%%%%%%%%%%%%%%%%%%%%%%%%%%%%%%%%%%%%%%%%%%%%%%%%
%%%%%%%%%%%%%%%%%%%%%%%%%%%%%%%%%%%%%%%%%%%%%%%%
%%%%%%%%%%%%%%%%%%%%%%%%%%%%%%%%%%%%%%%%%%%%%%%%
\section*{Acknowledgments}
%%%%%%%%%%%%%%%%%%%%%%%%%%%%%%%%%%%%%%%%%%%%%%%%
%%%%%%%%%%%%%%%%%%%%%%%%%%%%%%%%%%%%%%%%%%%%%%%%
%%%%%%%%%%%%%%%%%%%%%%%%%%%%%%%%%%%%%%%%%%%%%%%%

We thank the referee for their reading of this manuscript and insightful comments that have improved this work.
Images and partial analysis were made using the pynbody python package (\url{https://github.com/pynbody/pynbody} \citep{2013ascl.soft05002P}. Numerical computations were [in part] carried out on Cray XC30 at Center for Computational Astrophysics, National Astronomical Observatory of Japan and the GPC supercomputer at the SciNet HPC Consortium \citep{2010JPhCS.256a2026L}. SciNet is funded by: the Canada Foundation for Innovation under the auspices of Compute Canada; the Government of Ontario; Ontario Research Fund - Research Excellence; and the University of Toronto. We thank E. Tasker, K. Sorai and T. Okamoto for helpful discussion that aided in improving this work.

%%%%%%%%%%%%%%%%%%%%%%%%%%%%%%%%%%%%%%%%%%%%%%%%
%%%%%%%%%%%%%%%%%%%%%%%%%%%%%%%%%%%%%%%%%%%%%%%%
%%%%%%%%%%%%%%%%%%%%%%%%%%%%%%%%%%%%%%%%%%%%%%%%
\bibliographystyle{mnras}
\bibliography{GalBib}

%%%%%%%%%%%%%%%%%%%%%%%%%%%%%%%%%%%%%%%%%%%%%%%%
%%%%%%%%%%%%%%%%%%%%%%%%%%%%%%%%%%%%%%%%%%%%%%%%
%%%%%%%%%%%%%%%%%%%%%%%%%%%%%%%%%%%%%%%%%%%%%%%%
\bsp
\label{lastpage}
%%%%%%%%%%%%%%%%%%%%%%%%%%%%%%%%%%%%%%%%%%%%%%%%
%%%%%%%%%%%%%%%%%%%%%%%%%%%%%%%%%%%%%%%%%%%%%%%%
%%%%%%%%%%%%%%%%%%%%%%%%%%%%%%%%%%%%%%%%%%%%%%%%
\end{document}